\documentclass[journal,twoside]{IEEEtran}
\usepackage{cite,graphicx,subfigure,url,stfloats,amsmath,amssymb}
\usepackage{hyperref}
\usepackage{upgreek}
\usepackage{balance}
\usepackage[OT1]{fontenc}
\usepackage{enumerate}
\usepackage{color}

\begin{document}

\title{Closed-form Approximation for Performance Bound of Finite Blocklength Massive MIMO Transmission}

\author
{Xiaohu~You,~\IEEEmembership{Fellow,~IEEE,} Bin~Sheng,
Yongming~Huang,~\IEEEmembership{Senior~Member,~IEEE,}
Wei~Xu,~\IEEEmembership{Senior~Member,~IEEE,}
Chuan~Zhang,~\IEEEmembership{Senior~Member,~IEEE,}
Dongming~Wang,~\IEEEmembership{Senior~Member,~IEEE,}
Pengcheng~Zhu,~\IEEEmembership{Senior~Member,~IEEE,} and Chen~Ji
\thanks{This work is supported by the National R\&D Program of China under Grant 2020YFB1806603. \textit{(Corresponding Authors: Xiaohu You and Bin Sheng.)}}
\thanks{X. You, B. Sheng, Y. Huang, W. Xu, C. Zhang, D. Wang, and P. Zhu are with National Mobile Communications Research Laboratory, Southeast University, Nanjing 210096, China and Purple Mountain Laboratories, Nanjing 211111, China. (email: \{xhyu, sbdtt, huangym, wxu, chzhang, wangdm, p.zhu\}@seu.edu.cn).}
\thanks{C. Ji is with School of Information Science and Technology, Nantong University, Nantong 226019, China. (email: gwidjin@ntu.edu.cn)}
}

\markboth{Journal of \LaTeX\ Class Files,~Vol., No., 2022}%
{X. You \MakeLowercase{\textit{et al.}}: Closed-form Approximation for Performance Bound of Finite Blocklength Massive MIMO Transmission}

\maketitle

\begin{abstract}
It is supposed that ultra-reliable low latency communication
(uRLLC) would continue to evolve in the future sixth generation
(6G) network, to provide enhanced capability towards extreme
connectivity, with the aid of well established multiple-input
multiple-output (MIMO) technology. Since the latency constraint
can be represented equivalently by the blocklength of a codeword,
channel coding theory at a finite blocklength plays an important
role in theoretic analysis of uRLLC. Based on Polyanskiy's and
Yang's asymptotic results on maximal achievable rate, we first
derive the proximate closed-form expressions for the expectation
and variance of channel dispersion. Then, the upper bound of
average maximal achievable rate is obtained for massive MIMO
systems under ideal independent and identically distributed fading
channels. Since almost all the fundamental parameters, including
the spatial degree-of-freedom (DoF), are considered, this
expression can be viewed as a performance bound of the
spatiotemporal two-dimension channel coding to some extent.
Moreover, it is shown by simulation and analysis, as the DoF goes
to infinite, MIMO systems reveals a nature of deterministic
transmission, since the average maximal achievable coding rate per
antenna can be achieved at each transmission. In this case, the
inversely proportional law observed therein implies that the
blocklength in the time domain can be further shortened at the
expense of spatial DoF. This exchangeability of space and time, to
support a given coding rate, paves a solid and feasible road for
us to further reduce latency in 6G uRLLC.

\end{abstract}

\begin{keywords}
Channel coding, Mobile Communication, MIMO.
\end{keywords}

\section{Introduction}


\PARstart{I}n the beyond fifth-generation (B5G) mobile
communication network, ultra-reliable low latency communications
(uRLLC) are designed to support a plethora of mission-critical
applications, such as autonomous vehicles and virtual reality,
where seamless and reliable data transmission is the key to the
feasibility of service \cite{Sachs2018, Liu2022}. The primary
objective of uRLLC, as per the third-generation partnership
project (3GPP), is to reduce the latency down to $1\;{\rm{ms}}$
and simultaneously guaranteeing reliability higher than 99.999\%
\cite{3GPP2017}. On the one hand, driven by the increasingly
stringent requirements in the emerging application scenarios
\cite{Sutton2019}, further enhanced ability of uRLLC is expected
towards 6G ${\rm{TK\upmu}}$ extreme connectivity \cite{Xu2023},
i.e., the latency is reduced from the order of ${\rm{ms}}$ to
${\rm{\upmu s}}$. On the other hand, new applications such as
extended reality (XR) will dissolve the boundary between uRLLC and
enhanced mobile broadband (eMBB) \cite{Saad2019}. As a result,
while error control is conventionally achieved by implementing the
hybrid automatic repeat request (HARQ) mechanism, the strict
latency constraint in uRLLC excludes multiple retransmissions
therein. Additionally, potential new features, such as high data
rate, also pose considerable challenges to 6G uRLLC \cite{XuWei,
You2021, Zorzi1}.


Implementing uRLLC relies on short packet length and small
transmission time interval, which make channel coding more
challenging. Enabled by the spatial freedom provided by MIMO, a
spatiotemporal two-dimension (2D) channel coding scheme is
proposed in \cite{You2023}, which can flexibly balance the
transmission reliability and latency under different scenarios. In
the link layer, there are some innovative techniques to achieve
uRLLC, such as grant-free access, interface diversity, and radio
resource allocation. Specifically, three grant-free access
retransmission schemes including reactive, k-repetition, and
proactive retransmission were analyzed in \cite{Liu2021}. On the
other hand, novel computing frameworks and network architecture
techniques are attracting much attention. By optimizing task
offloading and resource allocation at edge nodes, MEC systems
shorten the delay and promise to support various mission-critical
applications \cite{She2019}. By serving each user with multiple
links from different nodes, multi-connectivity is regarded as an
effective approach to ensure high network availability and
reliability \cite{She2018}. To reduce control-plane latency in
handover procedure, anticipatory networks predict the mobility of
users according to their mobility pattern and reserve resources
proactively \cite{Bui2017}.

\subsection{Prior Arts}
All along, packet error probability $\varepsilon$, blocklength
(i.e., codeword size) $n$, and coding rate $R$ (the number of
information bits per complex symbol) are the three fundamental
metrics involved in communication systems. They are highly
correlated to each other and neither is dispensable, just like the
vertexes of a triangle \cite{Shannon1968,Gallager1968,You2020}. In
2010, building upon Dobrushin's and Strassen's asymptotic results,
Y. Polyanskiy, H. V. Poor, and S. Verdu presented a new framework
with tight bounds on $R$ as a function of $n$ and $\varepsilon$
under an additive white Gaussian noise (AWGN) channel
\cite{Polyanskiy2010}. The rationale behind Polyanskiy's approach
is that for finite $n$, the coding rate becomes a random variable
parameterized by channel capacity and channel dispersion, where
the channel dispersion is also a random variable introduced as a
rate penalty to characterize the impact of $n$. In the sequel, a
series of works have extended their results to other kinds of
point-to-point channels. In a single-input single-output (SISO),
stationary coherent fading channel with additive Gaussian noise,
\cite{Polyanskiy2011} obtained a convenient two-term expression
for the channel dispersion which is found to rely highly on the
dynamics of the fading process. In \cite{Yang2012}, the
non-asymptotic bounds of the maximum coding rate were presented
which is found not to be monotonic with respect to the channel
coherence time. Consequently, there exists a coherence time that
maximizes the coding rate over noncoherent Rayleigh block-fading
channels. Its normal approximation in the high SNR regime was then
presented in \cite{Lancho2020}, which provides a tractable formula
for us to perform further analysis. The throughput of spectrum
sharing networks with rate adaptation is studied in \cite{Zorzi3}
and the closed-form expression for the secondary user activation
probability is derived by using the existing results on finite
blocklength. Following the Polyanskiy's approach, a series of
works have extended this result in single antenna to the field of
multiple antennas. The maximal achievable rate of finite
blocklength in quasi-static MIMO channels, under mild conditions
on the fading distribution, has been provided recently by
Yang\emph{~et~al.} \cite{Yang2014}. In \cite{Zorzi2}, the
closed-form expressions for the message decoding probabilities as
well as the throughput, the expected delay, and the error
probability of HARQ setups were derived, for
single-input-multiple-output (SIMO) systems. Recently, Collins
\emph{~et~al.} in \cite{Collins2016} obtained a channel dispersion
formula for the MIMO block-fading channel where the most
interesting result is that the normalized dispersion decreases
with a growing number of receive antennas.

In practical wireless systems, coherent detection is widely used
which requires the receiver to know the channel state information
(CSI). Usually, CSI is acquired by sending training pilots at the
transmitter. Remarkable literatures have been devoted to address
channel estimation problems under the consideration of a finite
blocklength. In \cite{Ostman} and \cite{Lancho}, a rigorous
framework for characterizing the error probability in the uplink
and downlink of massive MIMO with finite blocklength was proposed.
Based on this framework, upper bounds of error probability are
given by using the mismatched scaled nearest-neighbor (SNN)
decoding rule, and the saddlepoint approximation was provided for
numerical evaluation. In \cite{Potter}, when the channel coherence
time exceeds the blocklength interval in a non-ergodic case, the
rates were proven to converge to the generalized mutual
information subject to an outage probability. Compared with normal
approximation, the saddlepoint approximation is observed to be
more accurate for the case where the target rate is not close to
the mean of the information density \cite{Lancho}\cite{Kislal}.
While in general cases, it could be difficult to achieve a concise
closed-form expression, when the saddlepoint approximation is
used.

\subsection{Main Contributions}
The concept of massive MIMO was first proposed by Marzetta in
2010, which envisions more antennas at transceiver than that of
conventional MIMO \cite{Marzetta2010}. So far, massive MIMO has
been adopted in 5G, as a key physical-layer technology to meet the
demand for higher data capacity for mobile networks. It is
supposed that the massive MIMO technique will continue to evolve
and more antennas are considered to employ, as the Tera Hertz
frequency band is envisioned for future 6G. More antennas would
provide more augmented DoFs which leads to some remarkable
properties, such as channel hardening and decorrelation.

In this paper, expectation and variance of the maximal achievable
rate in a massive MIMO system with finite blocklength are
investigated. The channel is assumed to be quasi-static so that
random fading coefficients remain constant over the duration of
each codeword. This is a typical assumption for uRLLC where the
blocklength is usually short enough. We assume no CSI available at
the transmitter partly because there can be insufficient time for
the receiver to feedback CSI in a frequency division duplex (FDD)
system. Moreover, isotropic codewords are assumed due to the lack
of CSI information. The normal approximation of maximal achievable
rate at finite blocklength is presented in \cite{Yang2014} for
MIMO sysyems. However, it is still inconvenient to use it to
analyze the impact of the numbers of antennas on the packet error
probability, because there is no closed-form formulae to describe
their relationships. To circumvent this difficulty, we therefore
make some approximations and turn to acquire the expectation and
variance of channel dispersion.

The main contributions of this paper are summarized as follows.

\begin{itemize}
    \item The closed-form expressions for expectation and variance of channel
dispersion in a massive MIMO system with independent and
identically distributed (i.i.d.) fading channels are derived.
Although the results are obtained under the assumption that the
number of antennas goes to infinity, it is observed through
numerical simulations that they are also accurate for a finite,
and even not-too-large, number of antennas case.
    \item  we show that, for a massive MIMO system with $M$ transmit
antennas and $N$ receive antennas, there exists an approximate
solution to the general relationship among latency $n$,
reliability $\varepsilon$, desired average data rate ${\bar R}$,
and spatial DoF $m$, where $m = \min \left\{ {M,N} \right\}$. The
embodiment of this relationship can be viewed as a performance
bound of the spatiotemporal 2D channel coding to some extent.
    \item In the high SNR regime, this 2D channel coding rate can be
    simplified to a pretty concise format as
\begin{equation}
\frac{{\bar R}}{m} \le {\log _2}\left( {1 + {\rho}} \right) -
\sqrt {\frac{1}{{mn}}} {\Phi ^{ - 1}}\left( \varepsilon  \right),
\end{equation}
where $\rho$ represents the signal-to-noise ratio (SNR) and ${\Phi
^{ - 1}}\left(  \cdot  \right)$ denotes the inverse of the
Gaussian $Q$-function
\begin{equation}
\Phi \left( x \right) \buildrel \Delta \over = \int_x^\infty
{\frac{1}{{\sqrt {2\pi } }}} {e^{{{ - {t^2}} \mathord{\left/
 {\vphantom {{ - {t^2}} 2}} \right.
 \kern-\nulldelimiterspace} 2}}}{\rm{d}}t.
\end{equation}
More interestingly, when the value of $m$ is large enough, it
turns out to be

\begin{equation}
n \approx \frac{{m {{\left[ {{\Phi ^{ - 1}}\left( \varepsilon
\right)} \right]}^2}}}{{{{\left[ {m{{\log }_2}\left( {1 + {\rho}}
\right) - \bar R} \right]}^2}}} \propto \frac{1}{m},
\end{equation}
which reveals an inversely proportional scaling law. That is, the
latency in the time domain can be reduced by increasing the DoFs
in the space domain, for the purpose of maintaining a ceratin
level of performance target.
\end{itemize}

The remainder of this paper is organized as follows. The system
model and relevant works are introduced in Section II. Section III
presents a complete derivation of the expectation and variance for
channel dispersion. The bounds on average maximal achievable rate
are investigated in Section IV. Section V provides numerical
results to reveal the potential relations of the most fundamental
parameters involved in MIMO communications.

{\emph{Notations}}: Boldface lower and upper case letters are used
to denote vectors and matrices, respectively. The notation
${\left( \cdot  \right)^T}$ and ${\left(  \cdot  \right)^H}$
denote the transpose and the conjugate transpose of a vector or
matrix, respectively. We use $\mathbf{I}_a$ to denote the identity
matrix of size $a \times a$. The mean, variance, and probability
of a random variable $x$ are illustrated by the operators
$\mathbb{E}\left\{ x \right\}$, ${\text{Var}}\left\{ x \right\}$,
and $\mathbb{P}\left\{ x \right\}$, respectively. The notation
${\cal C}{\cal N}\left( {0,{\sigma ^2}} \right)$ represents the
complex Gaussian distribution with zero mean and variance $\sigma
^2$ and ${\cal C}^{M \times N}$ denotes complex matrices with
dimension $M \times N$. Moreover, we use ${\text{tr}}\left(
{\mathbf{A}} \right)$ and ${\text{det}}\left( {\mathbf{A}}
\right)$ to denote the trace and determinant of matrix $\bf{A}$,
respectively. The Frobenius norm of a matrix $\bf{A}$ is denoted
by ${\left\| {\mathbf{A}} \right\|_{\text{F}}} \triangleq \sqrt
{{\text{tr}}\left( {{\mathbf{A}}{{\mathbf{A}}^H}} \right)}$. At
last, for two functions $f\left( x \right)$ and $g\left( x
\right)$, $f\left( x \right) = \mathcal{O}\left( {g\left( x
\right)} \right)$ means that $\lim {\sup _{x \to \infty }}\left|
{{{f\left( x \right)} \mathord{\left/
 {\vphantom {{f\left( x \right)} {g\left( x \right)}}} \right.
 \kern-\nulldelimiterspace} {g\left( x \right)}}} \right| < \infty
 $.

\section{System Model}
We consider an MIMO system with $M$ transmit antennas and $N$
receive antennas, operating over flat fading channels. Under the
assumption that the synchronization is perfect and the channel is
flat-fading quasi-static, the baseband equivalent discrete-time
input-output relationship can be written as

\begin{equation}
{\bf{Y}} = {\bf{HX}} + {\bf{W}},
\end{equation}
where ${\bf{X}} \in {{\cal C}^{M \times n}}$ is the signal
transmitted over $n$ time samples (channel uses), ${\bf{Y}} \in
{{\cal C}^{N \times n}}$ is the corresponding received signal, and
${\bf{H}} \in {{\cal C}^{N \times M}}$ contains the complex fading
coefficients, which are random but remain constant over the $n$
time samples. It should be noted that this assumption for channel
is usually reasonable for the case of uRLLC, due to the constrain
of latency. When an $i.i.d.$ Rayleigh fading channel is
considered, each entry of $\bf{H}$ is modelled as a Gaussian
variable with zero mean and unit variance. ${\bf{W}} \in {{\cal
C}^{N \times n}}$ denotes the additive noise at the receiver,
which is independent of ${\bf{H}}$ and also has $i.i.d.$ ${\cal
C}{\cal N}\left( {0,1} \right)$ entries.

According to \cite{Yang2014}, under the assumption of both
isotropic codewords and perfect CSI at the receiver, the normal
approximation to the maximal achievable rate is obtained as the
solution of
\begin{equation}
\varepsilon  = \mathbb{E}\left\{ {\Phi \left[ {\frac{{C\left(
{\mathbf{H}} \right) - {R^ * }\left( {n,\varepsilon }
\right)}}{{\sqrt {{{V\left( {\mathbf{H}} \right)} \mathord{\left/
 {\vphantom {{V\left( {\mathbf{H}} \right)} n}} \right.
 \kern-\nulldelimiterspace} n}} }}} \right]} \right\},
\end{equation}
where $n$ denotes the blocklength, $\varepsilon$ represents the
block error probability, and ${\Phi ^{ - 1}}\left(  \cdot \right)$
is the inverse of the Gaussian $Q$-function. $C\left( {\mathbf{H}}
\right)$ and $V\left( {\mathbf{H}} \right)$ denote the Shannon
capacity and channel dispersion conditioned on $\mathbf{H}$. When
the transmitter has no CSI, it has been proved in \cite{Abbe2013}
that the optimal power allocation should fulfill

\begin{equation}
{\mathbf{X}}{{\mathbf{X}}^H} = \frac{\rho }{M}{{\mathbf{I}}_M}.
\end{equation}
In this case, the capacity and dispersion are further expressed as
\begin{equation}
C\left( {\mathbf{H}} \right) = \sum\limits_{j = 1}^m {\log \left(
{1 + {{\rho {\lambda _j}} \mathord{\left/
 {\vphantom {{\rho {\lambda _j}} M}} \right.
 \kern-\nulldelimiterspace} M}} \right)}
\end{equation}
and
\begin{equation}
V\left( {\mathbf{H}} \right) = m - \sum\limits_{j = 1}^m
{\frac{1}{{{{\left( {1 + {{\rho {\lambda _j}} \mathord{\left/
 {\vphantom {{\rho {\lambda _j}} M}} \right.
 \kern-\nulldelimiterspace} M}} \right)}^2}}}},
\end{equation}
where $\lambda _j$ denotes the $j$-th eigenvalue of
${\mathbf{H}}{{\mathbf{H}}^H}$ for $M \geqslant N$ and
${{\mathbf{H}}^H}{\mathbf{H}}$ for $M<N$.



\section{Statistical Properties of Channel Dispersion}
Since the channel matrix is random, its fading dynamics on the
channel dispersion can be investigated by deriving its statistical
properties. Specifically, its expectation, together with the
capacity, can help us to analyze the average maximal achievable
rate. As most of the practical systems work in the positive SNR
regime, we consider in this paper only the case of ${\rho}>0$.

\subsection{Expectation}
Taking expectation on (8), we obtain
\begin{equation}
\begin{gathered}
  \mathbb{E}\left\{ {V\left( {\mathbf{H}} \right)} \right\} = \mathbb{E}\left\{ {m - \sum\limits_{j = 1}^m {\frac{1}{{{{\left( {1 + {{\rho {\lambda _j}} \mathord{\left/
 {\vphantom {{\rho {\lambda _j}} M}} \right.
 \kern-\nulldelimiterspace} M}} \right)}^2}}}} } \right\} \hfill \\
   = \mathbb{E}\left\{ {m - \sum\limits_{j = 1}^m {\frac{1}{{1 + {{2\rho {\lambda _j}} \mathord{\left/
 {\vphantom {{2\rho {\lambda _j}} M}} \right.
 \kern-\nulldelimiterspace} M} + {{\left( {{{\rho {\lambda _j}} \mathord{\left/
 {\vphantom {{\rho {\lambda _j}} M}} \right.
 \kern-\nulldelimiterspace} M}} \right)}^2}}}} } \right\} \hfill \\
   > \mathbb{E}\left\{ {m - \sum\limits_{j = 1}^m {\frac{1}{{{{2\rho {\lambda _j}} \mathord{\left/
 {\vphantom {{2\rho {\lambda _j}} M}} \right.
 \kern-\nulldelimiterspace} M} + {{\left( {{{\rho {\lambda _j}} \mathord{\left/
 {\vphantom {{\rho {\lambda _j}} M}} \right.
 \kern-\nulldelimiterspace} M}} \right)}^2}}}} } \right\} \hfill \\
   = m - \mathbb{E}\left\{ {\sum\limits_{i = 1}^m {\frac{{{{{M^2}} \mathord{\left/
 {\vphantom {{{M^2}} {{\rho ^2}}}} \right.
 \kern-\nulldelimiterspace} {{\rho ^2}}}}}{{{\lambda _i}\left( {{{2M} \mathord{\left/
 {\vphantom {{2M} {\rho  + {\lambda _i}}}} \right.
 \kern-\nulldelimiterspace} {\rho  + {\lambda _i}}}} \right)}}} } \right\} \hfill \\
   = m - \frac{M}{{2\rho }}\mathbb{E}\left\{ {\sum\limits_{i = 1}^m {\frac{1}{{{\lambda _i}}}} } \right\} + \frac{M}{{2\rho }}\mathbb{E}\left\{ {\sum\limits_{i = 1}^m {\frac{1}{{{{2M} \mathord{\left/
 {\vphantom {{2M} {\rho  + {\lambda _i}}}} \right.
 \kern-\nulldelimiterspace} {\rho  + {\lambda _i}}}}}} } \right\}. \hfill \\
\end{gathered}
\end{equation}
According to [30, Eq.~(41)], the first expectation on the
right-hand-side (RHS) of (9) is obtained directly. That is,

i) $M>N$
\begin{equation}
\begin{gathered}
  \mathbb{E}\left\{ {\sum\limits_{i = 1}^m {\left( {\frac{1}{{{\lambda _i}}}} \right)} } \right\} = \mathbb{E}\left\{ {{\text{tr}}\left[ {{{\left( {{\mathbf{H}}{{\mathbf{H}}^H}} \right)}^{ - 1}}} \right]} \right\} \hfill \\
  \quad \quad \quad \quad \quad  = \frac{N}{{M - N}}, \hfill \\
\end{gathered}
\end{equation}

ii) $M<N$
\begin{equation}
\begin{gathered}
  \mathbb{E}\left\{ {\sum\limits_{i = 1}^m {\left( {\frac{1}{{{\lambda _i}}}} \right)} } \right\} = \mathbb{E}\left\{ {{\text{tr}}\left[ {{{\left( {{{\mathbf{H}}^H}{\mathbf{H}}} \right)}^{ - 1}}} \right]} \right\} \hfill \\
  \quad \quad \quad \quad \quad  = \frac{M}{{N - M}}. \hfill \\
\end{gathered}
\end{equation}

The second expectation on the RHS of (9) can be derived by using
the Stieltjes transform. The Stieltjes transform approach can
solve most problems involved in random matrices, such as the
distribution functions of the empirical eigenvalues of large
random matrices \cite{Tulino2004, Couillet2011}. After some
mathematical manipulations in Appendix A, we obtain its
closed-form expression in (14) on the top of the next page.

From (10) and (11), we see that the sum of the reciprocals of
eigenvalues is the power of the inverse of correlation matrix.
However, it is somewhat striking that the expectation of this
power does not exist when $M=N$ \cite{Jungnickel2002}. For this,
we have to go back to the original expression of (9). Since $1 +
{{\rho {\lambda _j}} \mathord{\left/
 {\vphantom {{\rho {\lambda _j}} M}} \right.
 \kern-\nulldelimiterspace} M} > 1$ for each eigenvalue, (9) can be derived by

\begin{equation}
\begin{gathered}
  \mathbb{E}\left\{ {V\left( {\mathbf{H}} \right)} \right\} = \mathbb{E}\left\{ {m - \sum\limits_{j = 1}^m {\frac{1}{{{{\left( {1 + {{\rho {\lambda _j}} \mathord{\left/
 {\vphantom {{\rho {\lambda _j}} M}} \right.
 \kern-\nulldelimiterspace} M}} \right)}^2}}}} } \right\} \hfill \\
  \quad \quad \;\;\quad \;\; > \mathbb{E}\left\{ {m - \sum\limits_{j = 1}^m {\frac{1}{{1 + {{\rho {\lambda _j}} \mathord{\left/
 {\vphantom {{\rho {\lambda _j}} M}} \right.
 \kern-\nulldelimiterspace} M}}}} } \right\} \hfill \\
  \quad \quad \;\;\quad \;\; = m - \frac{M}{\rho }\mathbb{E}\left\{ {\sum\limits_{i = 1}^m {\frac{1}{{{M \mathord{\left/
 {\vphantom {M {\rho  + {\lambda _i}}}} \right.
 \kern-\nulldelimiterspace} {\rho  + {\lambda _i}}}}}} } \right\}. \hfill \\
\end{gathered}
\end{equation}
By using the Stieljes transform and Mar{\v{c}}enko-Pastur law, we
obtain

\begin{equation}
\mathbb{E}\left\{ {V\left( {\mathbf{H}} \right)} \right\} = N +
\frac{N}{{2\rho }} - \frac{{N\sqrt {1 + 4\rho } }}{{2\rho }}.
\end{equation}

Fig. 1 gives comparison results of the channel dispersion for
$N=M$. The analytical results are calculated according to (13).
Most interestingly, although these expectations are derived for
large numbers of antennas, they are also quite accurate for
small-to-moderate antenna numbers in the high SNR regime. For
notational simplicity, we therefore omit the limit operator  in
(14) on the top of next page.

\begin{figure}[h]
     \begin{centering}
       \includegraphics[scale=0.55]{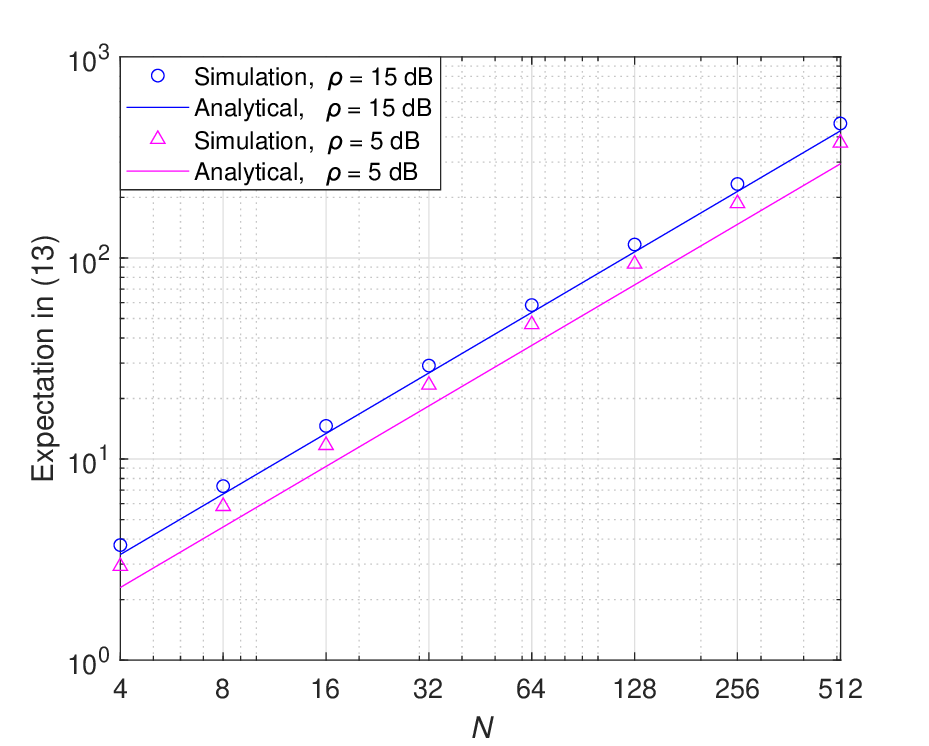}
       \caption{Expectation comparison for $N=M$.}
     \end{centering}
\end{figure}

\begin{figure*}[!t]
\begin{equation}
\label{Stieltjes} \mathbb{E}\left\{ {\sum\limits_{i = 1}^m
{\frac{1}{{{{2M} \mathord{\left/
 {\vphantom {{2M} {\rho  + {\lambda _i}}}} \right.
 \kern-\nulldelimiterspace} {\rho  + {\lambda _i}}}}}} } \right\} = \left\{ {\begin{array}{*{20}{c}}
  {\frac{N}{M}\left( {\frac{{\rho \left( {NM - {M^2}} \right)}}{{4{N^2}}} - \frac{M}{{2N}} + \frac{{\sqrt {{{\left( {\rho {M^2} - \rho MN + 2MN} \right)}^2} + 8\rho {N^2}{M^2}} }}{{4{N^2}}}} \right)\quad \quad N > M} \\
  {\frac{{\rho \left( {NM - {N^2}} \right)}}{{4{M^2}}} - \frac{N}{{2M}} + \frac{{\sqrt {{{\left( {\rho {N^2} - \rho MN + 2MN} \right)}^2} + 8\rho {N^2}{M^2}} }}{{4{M^2}}}\quad \quad \quad \quad N < M}
\end{array}} \right.
\end{equation}
\centering \rule[-10pt]{18cm}{0.05em}
\end{figure*}

Finally, by combining (10), (11), (13) and (14), we obtain the
complete close-form expression for the expectation of channel
dispersion. Most Interestingly, these expressions can be further
simplified to some constant in some special cases. For example, in
a massive MIMO system with $N=1$, as $M$ goes to infinity, we find
\begin{equation}
\mathop {\lim }\limits_{M \to \infty } \mathbb{E}\left\{
{\sum\limits_{i = 1}^m {\frac{1}{{{{2M} \mathord{\left/
 {\vphantom {{2M} {\rho  + {\lambda _i}}}} \right.
 \kern-\nulldelimiterspace} {\rho  + {\lambda _i}}}}}} } \right\} = 0
\end{equation}
and
\begin{equation}
\mathop {\lim }\limits_{M \to \infty } \frac{M}{{2\rho
}}\mathbb{E}\left\{ {\sum\limits_{i = 1}^m {\left(
{\frac{1}{{{\lambda _i}}}} \right)} } \right\} = \mathop {\lim
}\limits_{M \to \infty } \frac{M}{{2\rho }} \cdot \frac{1}{{M -
1}} = \frac{1}{{2\rho }}.
\end{equation}
In this case, the channel dispersion, for any positive SNR, can be
approximated by

\begin{equation}
\mathop {\lim }\limits_{M \to \infty } \mathbb{E}\left[ {V\left(
{\mathbf{H}} \right)} \right] \approx 1 - \frac{1}{{2\rho }}.
\end{equation}

\subsection{Variance}
The variance of channel dispersion is written by

\begin{equation}
\begin{gathered}
  \sigma _V^2 = \mathbb{E}\left\{ {{{\left( {V\left( {\mathbf{H}} \right) - \mathbb{E}\left[ {V\left( {\mathbf{H}} \right)} \right]} \right)}^2}} \right\} \hfill \\
  \quad \;\;\; = \mathbb{E}\left\{ {{{\left( {\sum\limits_{j = 1}^m {\frac{1}{{{{\left( {1 + {{\rho {\lambda _j}} \mathord{\left/
 {\vphantom {{\rho {\lambda _j}} {{N_t}}}} \right.
 \kern-\nulldelimiterspace} {{N_t}}}} \right)}^2}}}} } \right)}^2}} \right\} \hfill \\
  \quad \;\;\;\quad \;\;\;\quad \;\;\; - {\left[ {\mathbb{E}\left\{ {\sum\limits_{j = 1}^m {\frac{1}{{{{\left( {1 + {{\rho {\lambda _j}} \mathord{\left/
 {\vphantom {{\rho {\lambda _j}} {{N_t}}}} \right.
 \kern-\nulldelimiterspace} {{N_t}}}} \right)}^2}}}} } \right\}} \right]^2}. \hfill \\
\end{gathered}
\end{equation}
Since we already have the result of the second term of RHS in
(18), the remaining task is to obtain the closed-form formula of
the first term which has a denominator of order four.
Unfortunately, it is difficult to calculate the high order moments
of Wishart distributed matrices. Thus, we have to find some way to
reduce the order and try to make use of the available results of
their low order moments. Following this idea, the first term is
bounded as follows

\begin{equation}
\begin{gathered}
  \mathbb{E}\left\{ {{{\left( {\sum\limits_{j = 1}^m {\frac{1}{{{{\left( {1 + {{\rho {\lambda _j}} \mathord{\left/
 {\vphantom {{\rho {\lambda _j}} M}} \right.
 \kern-\nulldelimiterspace} M}} \right)}^2}}}} } \right)}^2}} \right\} \hfill \\
   = \mathbb{E}\left\{ {{{\left( {\sum\limits_{j = 1}^m {\frac{1}{{1 + 2{{\rho {\lambda _j}} \mathord{\left/
 {\vphantom {{\rho {\lambda _j}} M}} \right.
 \kern-\nulldelimiterspace} M} + {{\left( {{{\rho {\lambda _j}} \mathord{\left/
 {\vphantom {{\rho {\lambda _j}} M}} \right.
 \kern-\nulldelimiterspace} M}} \right)}^2}}}} } \right)}^2}} \right\} \hfill \\
   > \mathbb{E}\left\{ {{{\left( {\sum\limits_{j = 1}^m {\frac{1}{{2{{\rho {\lambda _j}} \mathord{\left/
 {\vphantom {{\rho {\lambda _j}} M}} \right.
 \kern-\nulldelimiterspace} M} + {{\left( {{{\rho {\lambda _j}} \mathord{\left/
 {\vphantom {{\rho {\lambda _j}} M}} \right.
 \kern-\nulldelimiterspace} M}} \right)}^2}}}} } \right)}^2}} \right\} \hfill \\
   = \mathbb{E}\left\{ {{{\left[ {\frac{M}{{2\rho }}\sum\limits_{i = 1}^m {\left( {\frac{1}{{{\lambda _i}}} - \frac{1}{{2{M \mathord{\left/
 {\vphantom {M {\rho  + {\lambda _i}}}} \right.
 \kern-\nulldelimiterspace} {\rho  + {\lambda _i}}}}}} \right)} } \right]}^2}} \right\} \hfill \\
   \triangleq {\text{G}}1 - 2 \cdot {\text{G2 + G3}} + {\text{G4,}} \hfill \\
\end{gathered}
\end{equation}
where we define

\begin{equation}
{\text{G}}1 \triangleq \mathbb{E}\left\{ {{{\left(
{\frac{M}{{2\rho }}} \right)}^2}\sum\limits_{i = 1}^m {{{\left(
{\frac{1}{{{\lambda _i}}}} \right)}^2}} } \right\}
\end{equation}

\begin{equation}
{\text{G}}2 \triangleq \mathbb{E}\left\{ {{{\left(
{\frac{M}{{2\rho }}} \right)}^2}\sum\limits_{i = 1}^m
{\sum\limits_{j = 1}^m {\frac{1}{{{\lambda _i}\left( {{{2M}
\mathord{\left/
 {\vphantom {{2M} {\rho  + {\lambda _j}}}} \right.
 \kern-\nulldelimiterspace} {\rho  + {\lambda _j}}}} \right)}}} } } \right\},
\end{equation}

\begin{equation}
{\text{G}}3 \triangleq \mathbb{E}\left\{ {{{\left(
{\frac{M}{{2\rho }}} \right)}^2}\sum\limits_{i = 1,i \ne j}^m
{\frac{1}{{{\lambda _i}{\lambda _j}}}} } \right\},
\end{equation}
and
\begin{equation}
\begin{gathered}
  {\text{G4}} \triangleq  \hfill \\
  \mathbb{E}\left\{ {{{\left( {\frac{M}{{2\rho }}} \right)}^2}\sum\limits_{i = 1}^m {\sum\limits_{j = 1}^m {\frac{1}{{\left( {2{M \mathord{\left/
 {\vphantom {M {\rho  + {\lambda _i}}}} \right.
 \kern-\nulldelimiterspace} {\rho  + {\lambda _i}}}} \right)\left( {2{M \mathord{\left/
 {\vphantom {M {\rho  + {\lambda _j}}}} \right.
 \kern-\nulldelimiterspace} {\rho  + {\lambda _j}}}} \right)}}} } } \right\}. \hfill \\
\end{gathered}
\end{equation}

Now the original expression has been expanded to the sum of
multiple parts which all have a lower order. Moreover, some of
them already have closed-form results. Based on the properties of
a Wishart Matrix introduced in [32, Lemma~2.10], for $M>N+1$, we
have

\begin{equation}
{\text{G}}1 = {\left( {\frac{M}{{2\rho }}}
\right)^2}\frac{{MN}}{{{{\left( {M - N} \right)}^3} - \left( {M -
N} \right)}},
\end{equation}
and
\begin{equation}
{\text{G}}3 = {\left( {\frac{M}{{2\rho }}}
\right)^2}\frac{{N\left( {N - 1} \right)}}{{\left( {M - N}
\right)\left( {M - N + 1} \right)}}.
\end{equation}

Unfortunately, it is difficult to derive the exact close-form
expressions for ${\text{G}}2$ and ${\text{G}}4$ by using
conventional integration approaches, since there is an additional
term ${{2M} \mathord{\left/
 {\vphantom {{2M} \rho }} \right.
 \kern-\nulldelimiterspace} \rho }$ appearing in the denominator. So, we resort to some
 approximations to the original expression. Otherwise, even if it can be solved, it may comprise of some special functions which makes the expression less tractable. By taking an independence assumption between two elements of ${\text{G}}2$, we obtain its closed-form result, for $M > N + 1$, as

\begin{equation}
\begin{gathered}
  {\text{G2}} = \zeta {\left( {\frac{M}{{2\rho }}} \right)^2}\frac{{{N^2}}}{{M - N}}\left[ {\frac{{N - M}}{{4\rho N}} - \frac{1}{2}} \right. \hfill \\
  \quad \quad \quad \quad \quad \quad \left. { + \frac{{\rho \sqrt {{{\left( {M - N + {{2N} \mathord{\left/
 {\vphantom {{2N} \rho }} \right.
 \kern-\nulldelimiterspace} \rho }} \right)}^2} + {{8{N^2}} \mathord{\left/
 {\vphantom {{8{N^2}} \rho }} \right.
 \kern-\nulldelimiterspace} \rho }} }}{{4N}}} \right], \hfill \\
\end{gathered}
 \end{equation}
where $\zeta  = {1 \mathord{\left/
 {\vphantom {1 {\left( {\psi N} \right)}}} \right.
 \kern-\nulldelimiterspace} {\left( {\psi N} \right)}}$ denotes
 the emendation parameter and $\psi $ is a real number which varies with $\rho$.

\emph{Proof}: See Appendix B.

Similar to the derivation of ${\text{G}}2$, we use the
independence assumption and express ${\text{G}}4$, for the case of
$M> N + 1$, as
\begin{equation}
\begin{gathered}
  {\text{G4}} = \xi \left[ {\frac{{M\left( {N - M} \right)}}{{8N}} - \frac{M}{{4\rho }}} \right. \hfill \\
  \quad \quad \quad \quad \quad \quad {\left. { + \frac{{M\sqrt {{{\left( {M - N + {{2N} \mathord{\left/
 {\vphantom {{2N} \rho }} \right.
 \kern-\nulldelimiterspace} \rho }} \right)}^2} + {{8{N^2}} \mathord{\left/
 {\vphantom {{8{N^2}} \rho }} \right.
 \kern-\nulldelimiterspace} \rho }} }}{{8N}}} \right]^2}, \hfill \\
\end{gathered}
\end{equation}
where $\xi$ is also an emendation parameter varying with $\rho$.

When $N=M$, the expectation term in (15) can be rewritten as

\begin{equation}
\begin{gathered}
  \mathbb{E}\left\{ {{{\left( {\sum\limits_{j = 1}^m {\frac{1}{{{{\left( {1 + {{\rho {\lambda _j}} \mathord{\left/
 {\vphantom {{\rho {\lambda _j}} M}} \right.
 \kern-\nulldelimiterspace} M}} \right)}^2}}}} } \right)}^2}} \right\} \hfill \\
   < \mathbb{E}\left\{ {{{\left( {\sum\limits_{j = 1}^m {\frac{1}{{1 + {{\rho {\lambda _j}} \mathord{\left/
 {\vphantom {{\rho {\lambda _j}} M}} \right.
 \kern-\nulldelimiterspace} M}}}} } \right)}^2}} \right\} \hfill \\
   = {\left( {\frac{M}{\rho }} \right)^2}\mathbb{E}\left\{ {\sum\limits_{i = 1}^m {\sum\limits_{j = 1}^m {\frac{1}{{\left( {{M \mathord{\left/
 {\vphantom {M {\rho  + {\lambda _i}}}} \right.
 \kern-\nulldelimiterspace} {\rho  + {\lambda _i}}}} \right)}} \cdot \frac{1}{{\left( {{M \mathord{\left/
 {\vphantom {M {\rho  + {\lambda _j}}}} \right.
 \kern-\nulldelimiterspace} {\rho  + {\lambda _j}}}} \right)}}} } } \right\}. \hfill \\
\end{gathered}
\end{equation}
Expression in (28) has a form similar to G4, except for the
scaling factor of two. By using the same mathematical tools and
the independence assumption, we have

\begin{equation}
\mathbb{E}\left\{ {{{\left( {\sum\limits_{j = 1}^m
{\frac{1}{{{{\left( {1 + {{\rho {\lambda _j}} \mathord{\left/
 {\vphantom {{\rho {\lambda _j}} M}} \right.
 \kern-\nulldelimiterspace} M}} \right)}^2}}}} } \right)}^2}} \right\} <  \frac{{{N^2}\xi}}{{4{\rho ^2}}}{\left( {\sqrt {1 + 4\rho }  - 1}
\right)^2},
\end{equation}
where $\xi$ denotes the emendation parameter varying with $\rho$.

Fig. 2 shows simulation results of (28) for $N=M$. The analytical
results are calculated according to (29). As we know, the ordered
eigenvalues in the random matrix are in fact correlated to each
other, which introduces approximation errors. Fortunately, these
approximation errors can be marginal by adjusting the emendation
parameters, as shown in Fig 2.

\begin{figure}[h]
     \begin{centering}
       \includegraphics[scale=0.6]{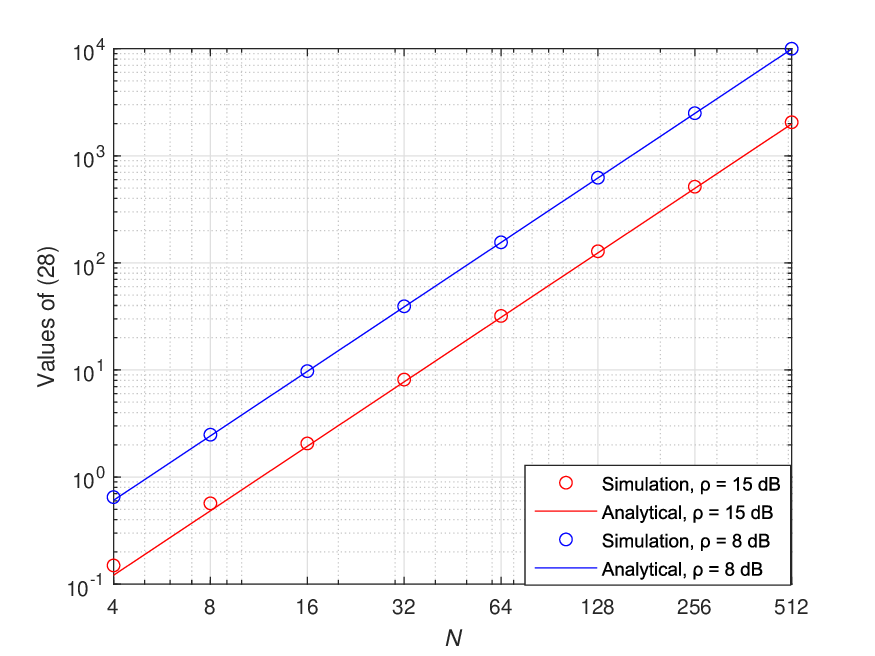}
       \caption{Variance comparison for $N=M$.}
     \end{centering}
\end{figure}

\section{Bounds on Average Maximal Achievable Rate}
Since the variance of channel dispersion decreases largely as $N$
goes to infinite, it can be treated as a deterministic number to
some extent. In this case for small $n$, we have
\begin{equation}
\begin{gathered}
  \varepsilon  > \Phi \left[ {\mathbb{E}\left\{ {\frac{{C\left( {\mathbf{H}} \right) - {R^ * }\left( {n,\varepsilon } \right)}}{{\sqrt {{{V\left( {\mathbf{H}} \right)} \mathord{\left/
 {\vphantom {{V\left( {\mathbf{H}} \right)} n}} \right.
 \kern-\nulldelimiterspace} n}} }}} \right\}} \right] \hfill \\
  \quad \quad  = \Phi \left[ {\frac{{\mathbb{E}\left\{ {C\left( {\mathbf{H}} \right)} \right\} - {R^ * }\left( {n,\varepsilon } \right)}}{{\mathbb{E}\left\{ {\sqrt {{{V\left( {\mathbf{H}} \right)} \mathord{\left/
 {\vphantom {{V\left( {\mathbf{H}} \right)} n}} \right.
 \kern-\nulldelimiterspace} n}} } \right\}}}} \right]. \hfill \\
\end{gathered}
\end{equation}
As a result, the upper bound of the maximal average achievable
rate can be written as
\begin{equation}
{R^ * }\left( {n,\varepsilon } \right) < \bar R = \mathbb{E}\left[
{C\left( {\mathbf{H}} \right)} \right] - \sqrt
{\frac{{\mathbb{E}\left[ {V\left( {\mathbf{H}} \right)}
\right]}}{n}} {\Phi ^{ - 1}}\left( \varepsilon  \right),
\end{equation}
where the expectation of channel dispersion can be found in (10),
(11), (13), and (14). It should be noted that the big $O$ term in
(31) is neglected, since it has a relatively much smaller value,
when compared with the capacity of massive MIMO systems.

Although these expressions are accurate, they seem to be so
complicated that it is not convenient to use them in performance
analysis. In fact, the results with concise expressions are more
useful in studying the relationship between the blocklength and
coding rate.

\subsection{High SNR Regime}
In the high per-antenna SNR  regime, i.e., ${\rho \mathord{\left/
 {\vphantom {\rho  M}} \right.
 \kern-\nulldelimiterspace} M} \gg 1$, the
channel dispersion can be approximated as

\begin{equation}
\begin{gathered}
  \mathbb{E}\left\{ {V\left( {\mathbf{H}} \right)} \right\} = m - \mathbb{E}\left\{ {\sum\limits_{j = 1}^m {\frac{1}{{{{\left( {1 + {{\rho {\lambda _j}} \mathord{\left/
 {\vphantom {{\rho {\lambda _j}} M}} \right.
 \kern-\nulldelimiterspace} M}} \right)}^2}}}} } \right\} \hfill \\
  \quad \quad \quad \quad  \approx m - \frac{{{M^2}}}{{{\rho ^2}}}\mathbb{E}\left\{ {\sum\limits_{j = 1}^m {\frac{1}{{\lambda _j^2}}} } \right\}. \hfill
  \\
\end{gathered}
\end{equation}
Based on the properties of Wishart matrix given in [32,
Lemma~2.10], we obtain

\begin{equation}
\mathbb{E}\left[ {V\left( {\mathbf{H}} \right)} \right] \approx m
- \frac{{{M^2}}}{{{\rho ^2}}} \cdot \frac{{NM}}{{{{\left( {M - N}
\right)}^3} - \left( {M - N} \right)}}.
\end{equation}
Interestingly, as the SNR per antenna goes to infinity, (33) is be
further simplified approximately to

\begin{equation}
\mathbb{E}\left[ {V\left( {\mathbf{H}} \right)} \right] \approx m.
\end{equation}
In some cases, if one of $M$ and $N$ is much larger than the
other, (34) holds apparently for  ${\rho \mathord{\left/
 {\vphantom {\rho  M}} \right.
 \kern-\nulldelimiterspace} M} \gg 1$.

\emph{Proof}: See Appendix C.

On the other hand, the capacity in the high SNR regime also has a
simple approximated formula as \cite{Zheng2003}

\begin{equation}
\mathbb{E}\left[ {C\left( {\mathbf{H}} \right)} \right] \approx
m\log \left( {1 + \rho } \right).
\end{equation}
Consequently, a concise and analytically tractable expression for
the average coding rate at a finite blocklength is achieved as

\begin{equation}
\bar R \approx m\log \left( {1 + \rho } \right) - \sqrt
{\frac{m}{n}} {\Phi ^{ - 1}}\left( \varepsilon  \right).
\end{equation}

\subsection{Normalized Maximal Achievable Rate}

As we know, after performing singular value decomposition (SVD) or
eigenvalue decomposition, the MIMO channel can be transformed into
multiple parallel orthogonal links. As a result, the average
maximal achievable rate over all the links plays an important role
in the performance analysis of massive MIMO systems at a finite
blocklength. Following this idea, we divide (36) by $m$ and obtain
\begin{equation}
\frac{{\bar R}}{m} \approx {\log _2}\left( {1 + \rho } \right) -
\sqrt {\frac{1}{{mn}}} {\Phi ^{ - 1}}\left( \varepsilon  \right),
\end{equation}
which can also be viewed as the rate bound of spatiotemporal 2D
channel coding. Although the potential relations of the most
fundamental parameters involved in MIMO communications has been
revealed in many existing results, it is the first time to
describe them in a concise and accurate closed-form expression. To
gain a deep insight into the latency, we focus on $n$ and $m$ in
the following analysis and let $\rho$ and $\epsilon$ be any real
positive numbers. Moreover, the channel matrix is assumed to be a
square matrix with full rank, for the simplicity of analysis. In
this case, from (37), we notice that the average data rate per Hz
per antenna remains unchanged, when we fix the value of $mn$. This
fact implies that $n$ and $m$ can be thought of being reciprocal
to some extent and thus exchanging $n$ by $m$ is possible in
theory. In conventional MIMO systems, channel coding is conducted
in the time domain on each link independently. So, if $n$
decreases, the reliability can not be sustained any more, for a
given coding rate. However, the observed reciprocal phenomenon
suggests us that we can perform channel coding in the space domain
along the links and exchange $n$ by $m$ to keep the total
performance unchanged. Indeed, when the value of $m$ is large
enough, we obtain
\begin{equation}
n \approx \frac{{m {{\left[ {{\Phi ^{ - 1}}\left( \varepsilon
\right)} \right]}^2}}}{{{{\left[ {m{{\log }_2}\left( {1 + \rho }
\right) - \bar R} \right]}^2}}} \propto \frac{1}{m},
\end{equation}
which illustrates that latency is inversely proportional to the
spatial DoF. This scaling law is just a basic principle behind the
2D channel coding presented in \cite{You2023}. (38) also provides
an upper bound for its maximal coding rate. More interestingly, it
can be supposed that for an extreme case of $n=1$, when $m$ goes
to infinity, the date rate per Hz per antenna, i.e. ${{\bar R}
\mathord{\left/
 {\vphantom {{\bar R} m}} \right.
 \kern-\nulldelimiterspace} m}$, approaches the Shannon capacity in AWGN channel.


\section{Numerical Results and Analysis}
Computer simulations were carried out in this section to verify
the analytical results. We check the expectation of channel
dispersion at first and then perform numerical results to study
maximal achievable rate. Finally, the impact of imperfect CSI on
the rate is investigated. In all the simulations, we set $c = {N
\mathord{\left/ {\vphantom {N M}} \right.
 \kern-\nulldelimiterspace} M}$.

\subsection{Statistical Properties of Channel Dispersion}
Fig. 3 gives numerical evaluation under different antenna
configurations where the analytical results are calculated
according to (10), (11), (13), and (14). It should be noted that
in Fig. 3, although the analytical results coincide well with the
simulation curves, there is still a visible small gap. The reason
is that an ignorable term is omitted in the denominator of (9).
Most strikingly, we find that (39) is accurate for all the
dimensions of the matrix in the test. Figs.~4 and 5 give the
simulation results to evaluate its accuracy. From both figures,
the small gap diminishes by using (39).

\begin{figure}[h]
     \begin{centering}
       \includegraphics[scale=0.6]{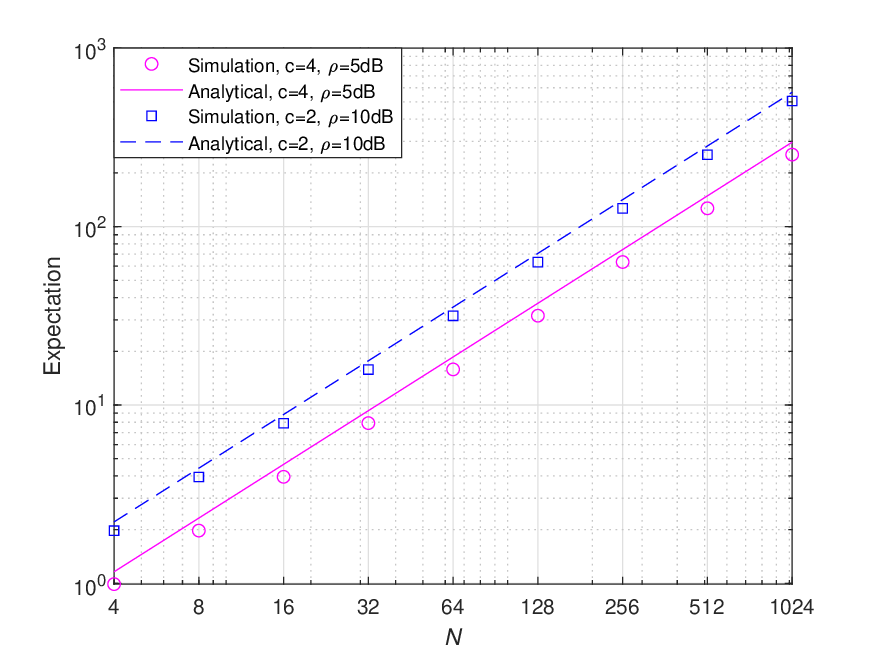}
       \caption{Expectation of channel dispersion for $N>M$.}
     \end{centering}
\end{figure}

\begin{figure*}[!t]
\begin{equation}
\mathbb{E}\left\{ {V\left( {\mathbf{H}} \right)} \right\} =
\left\{ {\begin{array}{*{20}{c}}
  {M - \frac{{{M^2}}}{{2\rho \left( {N - M} \right)}} + \frac{{{M^2}}}{{2\rho N}}\left( {\frac{{\rho \left( {NM - {N^2}} \right)}}{{4{M^2}}} - \frac{N}{{2M}} + \frac{{\sqrt {{{\left( {\rho {N^2} - \rho MN + 2MN} \right)}^2} + 8\rho {N^2}{M^2}} }}{{4{M^2}}}} \right)\quad \quad N > M} \\
  {\quad \quad \quad \quad \quad \quad \quad \quad \quad \quad \quad \quad N + \frac{N}{{2\rho }} - \frac{{N\sqrt {1 + 4\rho } }}{{2\rho }}\quad \quad \quad \quad \quad \quad \quad \quad \quad \quad \quad \quad \quad \quad N = M} \\
  {N - \frac{{MN}}{{2\rho \left( {M - N} \right)}} + \frac{N}{{2\rho }}\left( {\frac{{\rho \left( {NM - {M^2}} \right)}}{{4{N^2}}} - \frac{M}{{2N}} + \frac{{\sqrt {{{\left( {\rho {M^2} - \rho MN + 2MN} \right)}^2} + 8\rho {N^2}{M^2}} }}{{4{N^2}}}} \right)\quad \quad \quad N < M}
\end{array}} \right.
\end{equation}
\centering \rule[-10pt]{18cm}{0.05em}
\end{figure*}

\begin{figure}[h]
     \begin{centering}
       \includegraphics[scale=0.5]{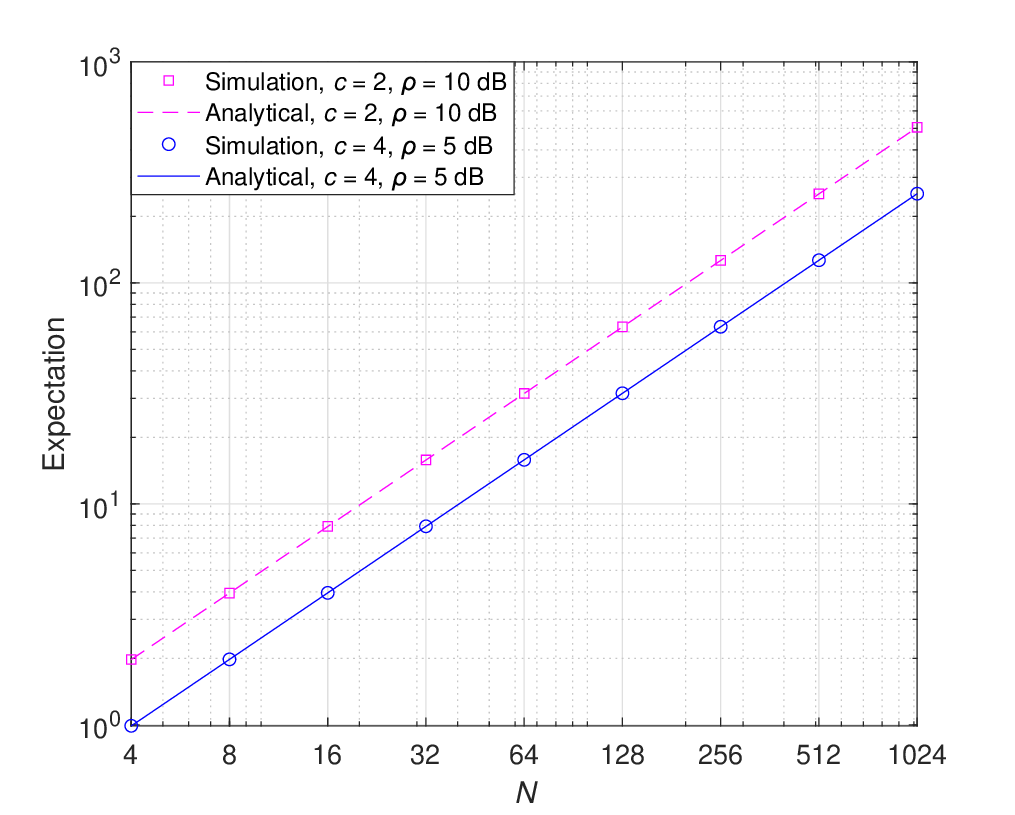}
       \caption{Expectation of channel dispersion for $N>M$.}
     \end{centering}
\end{figure}

\begin{figure}[h]
     \begin{centering}
       \includegraphics[scale=0.6]{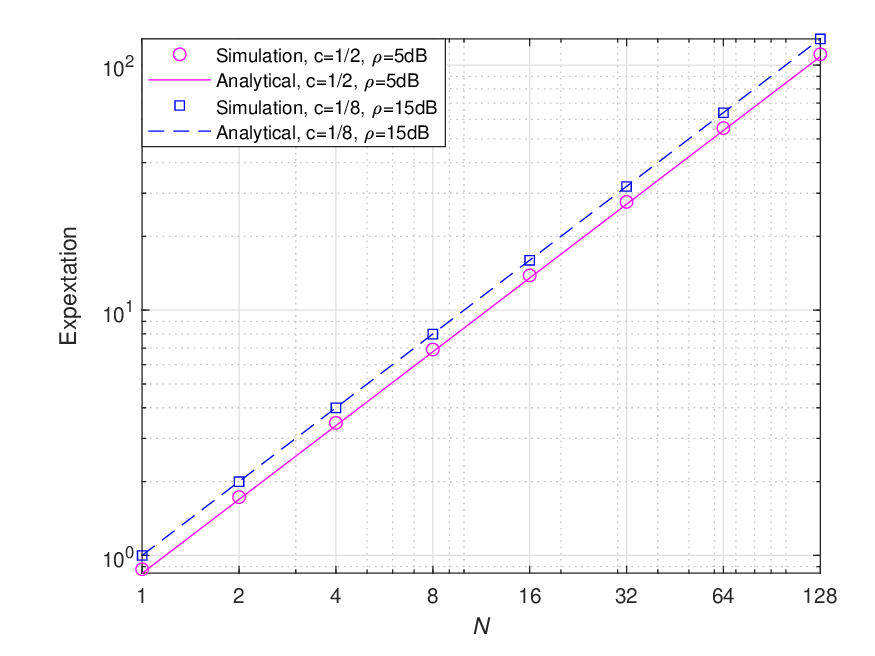}
       \caption{Expectation of channel dispersion for $N<M$.}
     \end{centering}
\end{figure}

Then, we perform simulations to investigate the variance of
channel dispersion. Fig. 6 illustrates the simulation results for
different antenna configurations where the analytical results are
calculated according to (24)-(27). The emendation parameters are
set to $\psi {\text{ = }}{\text{57.5}}$ and $\xi {\text{ =
0}}{\text{.5}}$ for $\rho  = 5\,{\text{dB}}$ and $\psi {\text{ =
}}{\text{370}}$ and $\xi {\text{ = 0}}{\text{.6}}$ for $\rho  =
7\,{\text{dB}}$. As can be seen from Fig.~6 that the variance of
channel dispersion is expected to be small, when compared with the
expectation, which highlights the dominant role of expectation in
the analysis of the system performance .

\begin{figure}[h]
     \begin{centering}
       \includegraphics[scale=0.63]{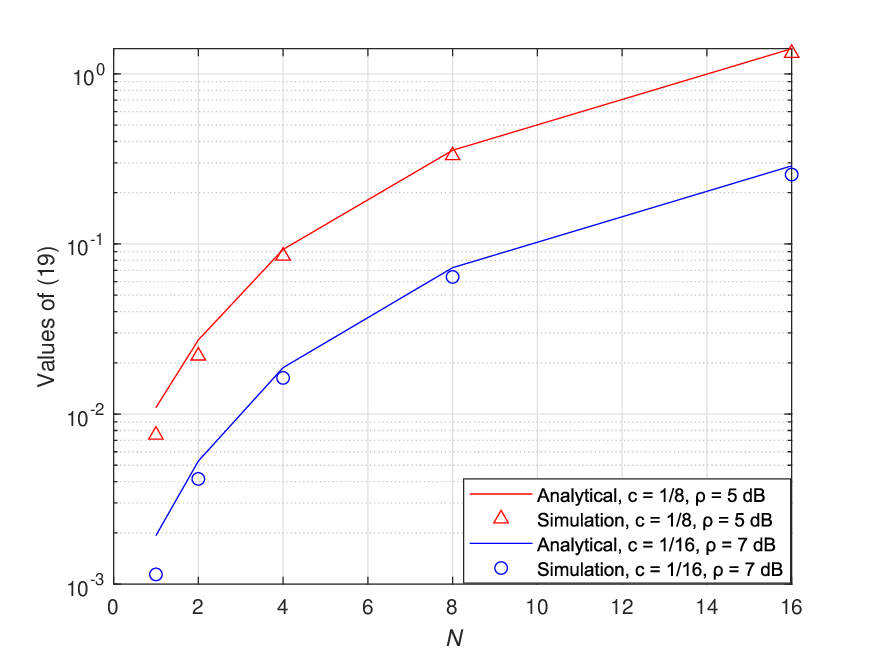}
       \caption{Comparison of analytical and simulation results.}
     \end{centering}
\end{figure}

\subsection{Normal Approximation}

Fig. 7 shows the block error probability for the case of $\rho =
20\, {\text{dB}}$ and $c = 8$. For normalization, the maximal
achievable rate is set to $6.5N$ bit/s/Hz, where $N$ denotes the
number of receive antennas. Simulation results are obtained by
using (5), while the analytical results are calculated according
to (30). As $n$ grows to 50, the gap between simulation and
analytical results becomes noticeable, even though the trends of
these curves remain the same. This gap mainly comes from that
channel capacity ${C\left( {\mathbf{H}} \right)}$ which has a
non-negligible variance. Fortunately, this offset can be fixed by
subtracting a standard deviation of the channel capacity from
(30). That is, the approximation for $n=50$, can be rewritten by

\begin{equation}
\varepsilon  \approx \Phi \left[ {\frac{{\mathbb{E}\left\{
{C\left( {\mathbf{H}} \right)} \right\} - {\sigma _C} - {R^ *
}\left( {n,\varepsilon } \right)}}{{\mathbb{E}\left\{ {\sqrt
{{{V\left( {\mathbf{H}} \right)} \mathord{\left/
 {\vphantom {{V\left( {\mathbf{H}} \right)} n}} \right.
 \kern-\nulldelimiterspace} n}} } \right\}}}} \right],
\end{equation}
where $\sigma _C$ denotes the standard deviation. According to
\cite{Couillet2011}, the channel capacity in our channel model
obeys the Gaussian distribution with variance

\begin{equation}
\sigma _C^2 =  - \log \left( {1 - \frac{{c \cdot {\mu _F}{{\left(
z \right)}^2}}}{{{{\left[ {1 + c \cdot {\mu _F}\left( z \right)}
\right]}^2}}}} \right),
\end{equation}
where
\begin{equation}
{\mu _F}\left( z \right) = \frac{{1 - c}}{{2cz}} - \frac{1}{{2c}}
- \frac{{\sqrt {{{\left( {1 - c - z} \right)}^2} - 4cz} }}{{2cz}}
\end{equation}
and
\begin{equation}
z =  - \frac{1}{\rho }.
\end{equation}
As can be seen from Fig. 7, the analytical results match well with
the numerical results especially in the regime of large $N$, when
using (40) for $n=50$.

\begin{figure}[h]
     \begin{centering}
       \includegraphics[scale=0.53]{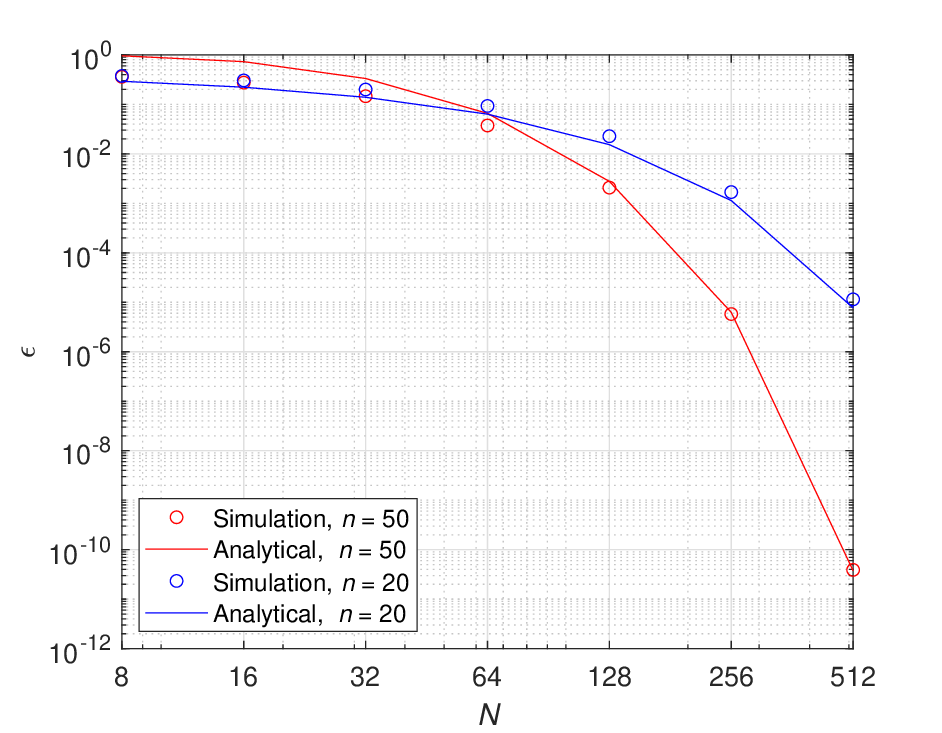}
       \caption{Comparison of block error probability.}
     \end{centering}
\end{figure}

\subsection{Bounds on Average Maximal Achievable Rate}
This subsection presents numerical results to depict the potential
relationships among blocklength, error probability, coding rate,
and spatial DoF, based on (31). Fig. 8 shows the average maximal
achievable rate as a function of $n$ and $\varepsilon$ under
$\rho=15\,\rm{dB}$. Different values of $m$ are set to generate
multiple surfaces for comparison. Form Fig. 8, we can see that as
$m$ increases, $n$ becomes smaller for a given average rate and
error probability, which verifies that the latency in the time
domain can be compensated by the DoFs in the space domain.

\begin{figure}[h]
     \begin{centering}
       \includegraphics[scale=0.55]{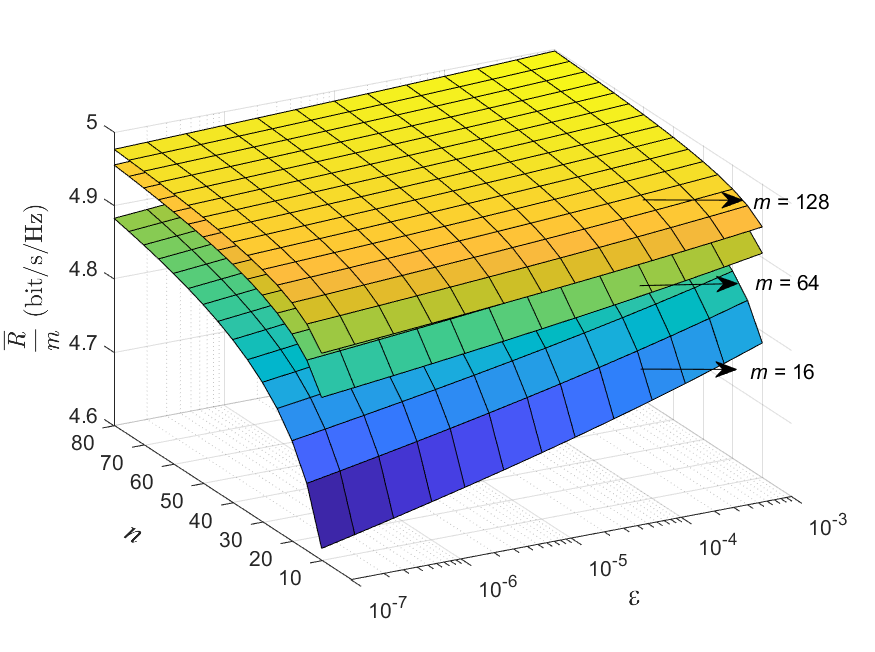}
       \caption{Comparison of average rate for different DoFs.}
     \end{centering}
\end{figure}
In Fig. 9, we fix the error probability at $\epsilon=10^{-6}$ and
vary $n$ to compare their rates. In the tests, the SNR is set to
$\rho=15\,\rm{dB}$. From Fig. 9, all the curves approach the line
of ${\log _2}\left( {1 + \rho } \right)$ as $m$ increases and
convergence speed accelerates as $n$ grows. It is expected that
when $m$ goes to infinity, each link of the MIMO system can
achieve the same Shannon capacity as that in an AWGN channel, for
any positive blocklength.

\begin{figure}[h]
     \begin{centering}
       \includegraphics[scale=0.65]{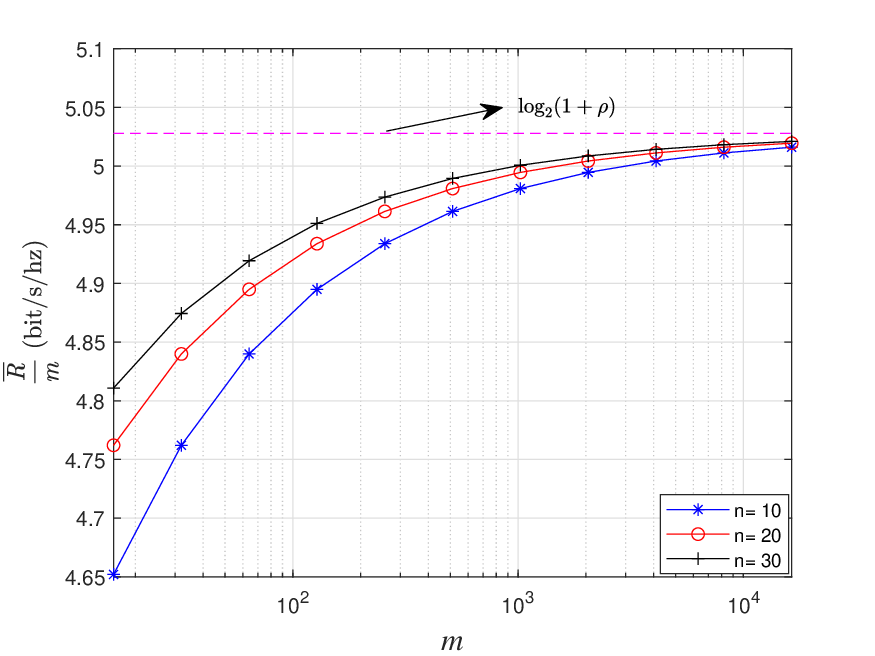}
       \caption{Comparison of average rate for different values of $n$.}
     \end{centering}
\end{figure}

In Fig. 10, the variance of ${{\bar R} \mathord{\left/
 {\vphantom {{\bar R} m}} \right.
 \kern-\nulldelimiterspace} m}$ is shown for $\epsilon=10^{-6}$
and $n=40$. From Fig. 10, we find that as the numbers of antennas
at both transmitter and receiver increase, the variance of
normalized average maximal achievable rate decreases quickly,
analogous to the channel hardening phenomena in a massive MIMO
system. That is, as the DoF goes to infinite, the massive MIMO
also reveals a feature of deterministic transmission, since the
average maximal achievable coding rate per antenna is achieved at
each transmission.

\begin{figure}[h]
     \begin{centering}
       \includegraphics[scale=0.45]{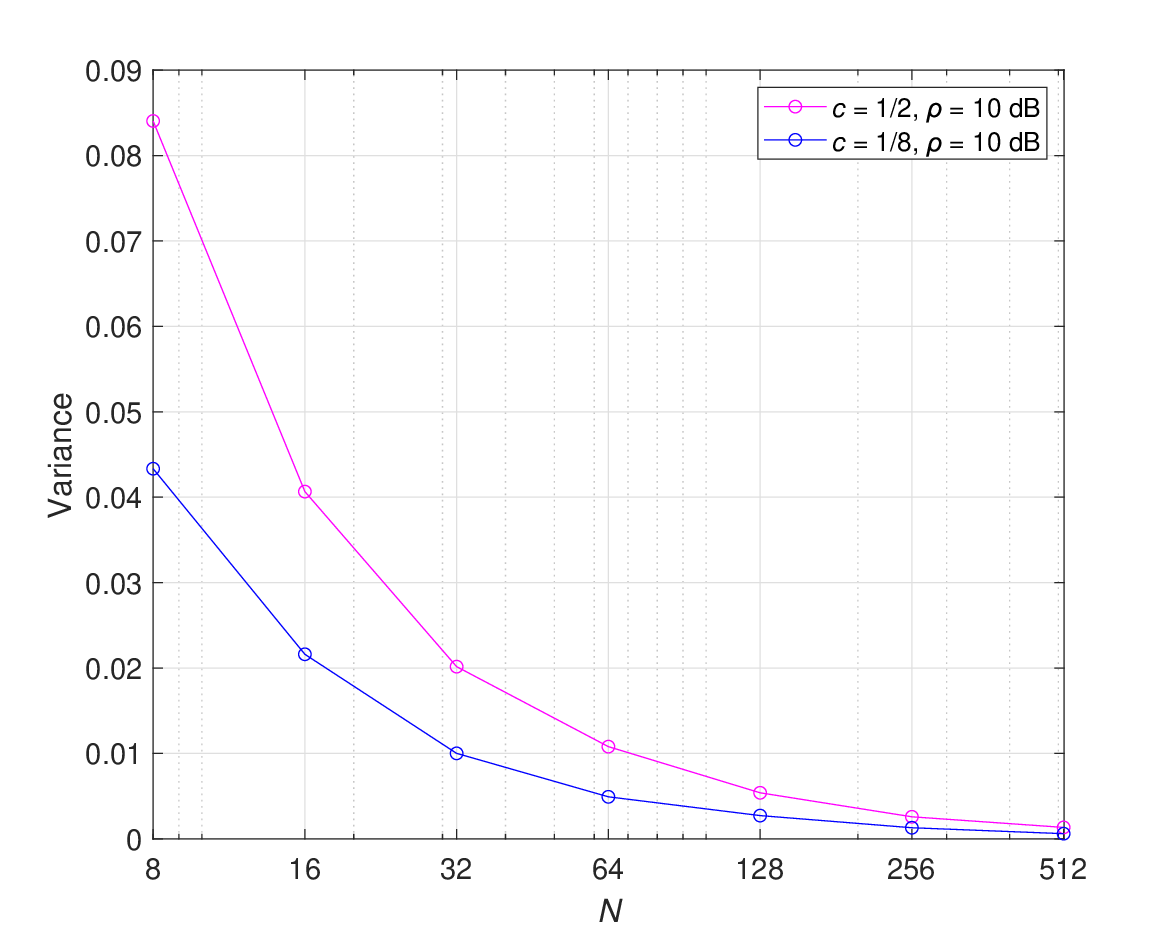}
       \caption{Variance of average rate for different values of $c$.}
     \end{centering}
\end{figure}

Moreover, we investigate the performance of error probability, as
plotted in Fig.~11 with respective to $m$ and $n$. With
$\rho=15\,\rm{dB}$ and ${{\bar R} \mathord{\left/ {\vphantom
{{\bar R} m}} \right. \kern-\nulldelimiterspace} m} = 4.95$. Form
Fig. 11, it suggests that we can either increase the blocklength
or use more antennas at both transmitter and receiver to maintain
a required reliability for a given rate and SNR.

\begin{figure}[h]
     \begin{centering}
       \includegraphics[scale=0.6]{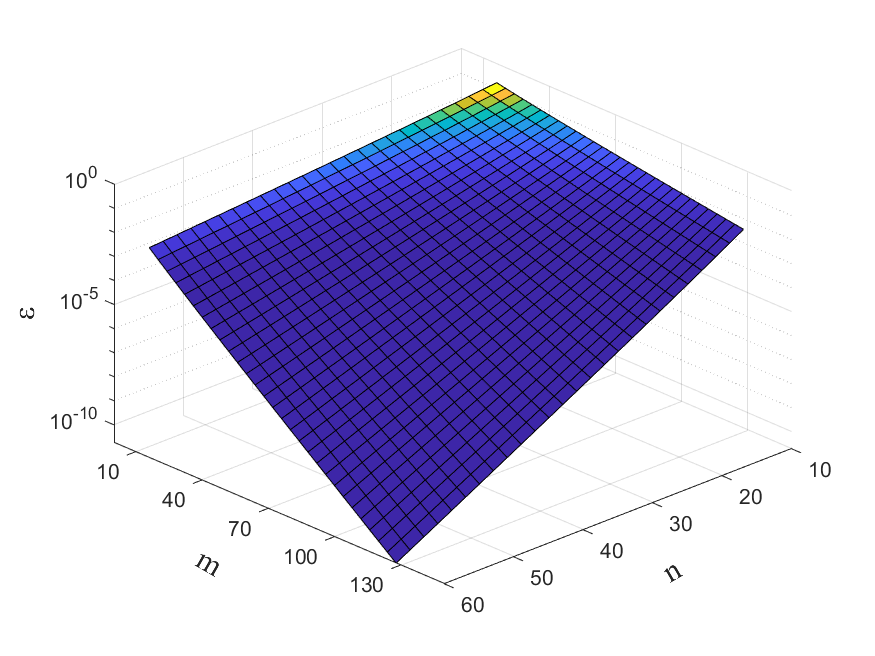}
       \caption{Comparison of blocklength for different DoFs.}
     \end{centering}
\end{figure}

\subsection{Imperfect CSI}

In practice, the CSI can be acquired by sending training symbols
at transmitter. The training overhead is a function of the channel
dynamics, because the faster the channel varies, the more training
symbols are needed in order to estimate the channel accurately.
According to \cite{Yang2012} and \cite{Lapidoth2005}, one way to
determine the training overhead is to estimate the capacity
penalty due to the lack of CSI. This makes it necessary to study
the capacity in a noncoherent setting, where neither the
transmitter nor the receiver are assumed to have the knowledge of
CSI. The loss of capacity, when compared with the coherent
setting, can be thought of as the minimum training overhead or
cost of CSI at receiver. Interestingly, it has been reported in
\cite{Hochwald2000, Zheng2002, Durisi2011} that as SNR goes to
infinity, the asymptotic ratio between the capacity and the
logarithm of SNR approaches $1 - {1 \mathord{\left/
 {\vphantom {1 T}} \right.
 \kern-\nulldelimiterspace} T}$, where $T$ denotes the coherence time of
the channel. It implies that the capacity penalty due to the lack
of CSI at receiver diminishes when the channel coherence time is
sufficiently large. In most practical systems, CSI is
indispensable for coherent demodulation. When the mobility is
relatively low, such as robots and autonomous vehicles in factory
automation \cite{Jayaweera}, there is usually enough time to send
pilots or training sequences. After the CSI is acquired, the
following blocks within a period of coherence time can be used to
carry data information. Even if the mobility is not low, some
advanced channel estimation methods can still be used to predict
the CSI, since the channel varies continuously in real
environments \cite{Yang}.

When the situation of imperfect CSI is considered, the impact
introduced by the channel estimate error is equivalently treated
as an additional noise. In this case, its variance can absorbed
into the thermal noise, which reduces the SNR at the receiver,
when compared with the perfect CSI scenario. Fig.~12 gives the
simulation results of (31) with $\varepsilon  = {10^{ - 5}}$ and
$n = 40$. From Fig. 12, as the variance of channel estimate error
decreases, the loss in average maximal achievable rate becomes
negligible. It should be noted that the overhead of training
symbols is not considered directly in the derived expressions,
because it mainly depends on the channel coherence time. If the
coherence time is large enough, the rate penalty due to the
transmission of known symbols at the receiver is expected to be
ignorable \cite{Tse2005}.
\begin{figure}[h]
     \begin{centering}
       \includegraphics[scale=0.45]{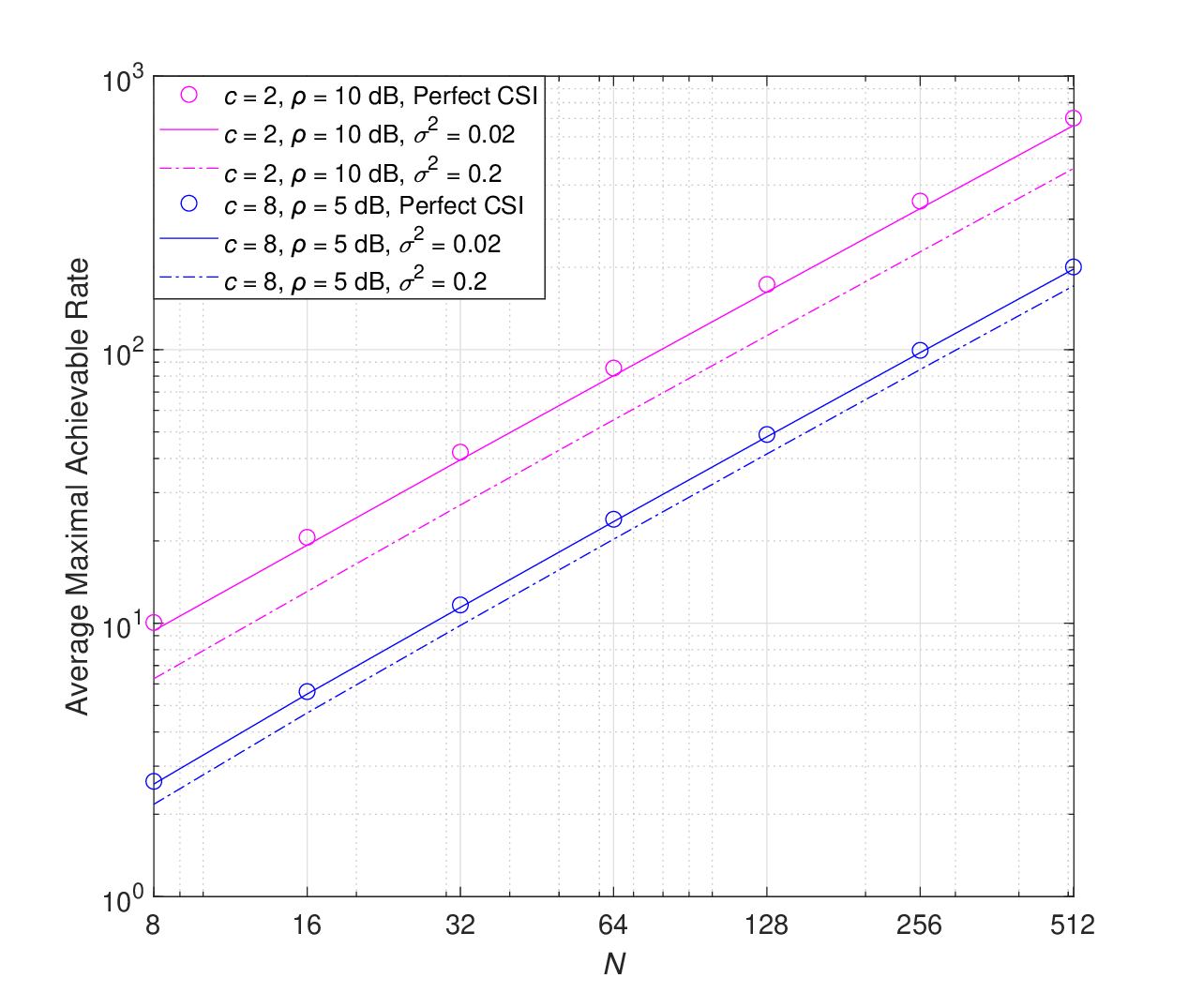}
       \caption{Performance comparison with imperfect CSI.}
     \end{centering}
\end{figure}

In Fig. 13, the variance of the maximal achievable rate in (31) is
tested by using the same parameters. Whether there is an channel
estimation error or not, the variance of rate becomes smaller, as
the number of antennas increases. In other words, the rate in each
transmission approaches its expectation, just like the effect of
channel hardening.
\begin{figure}[h]
     \begin{centering}
       \includegraphics[scale=0.5]{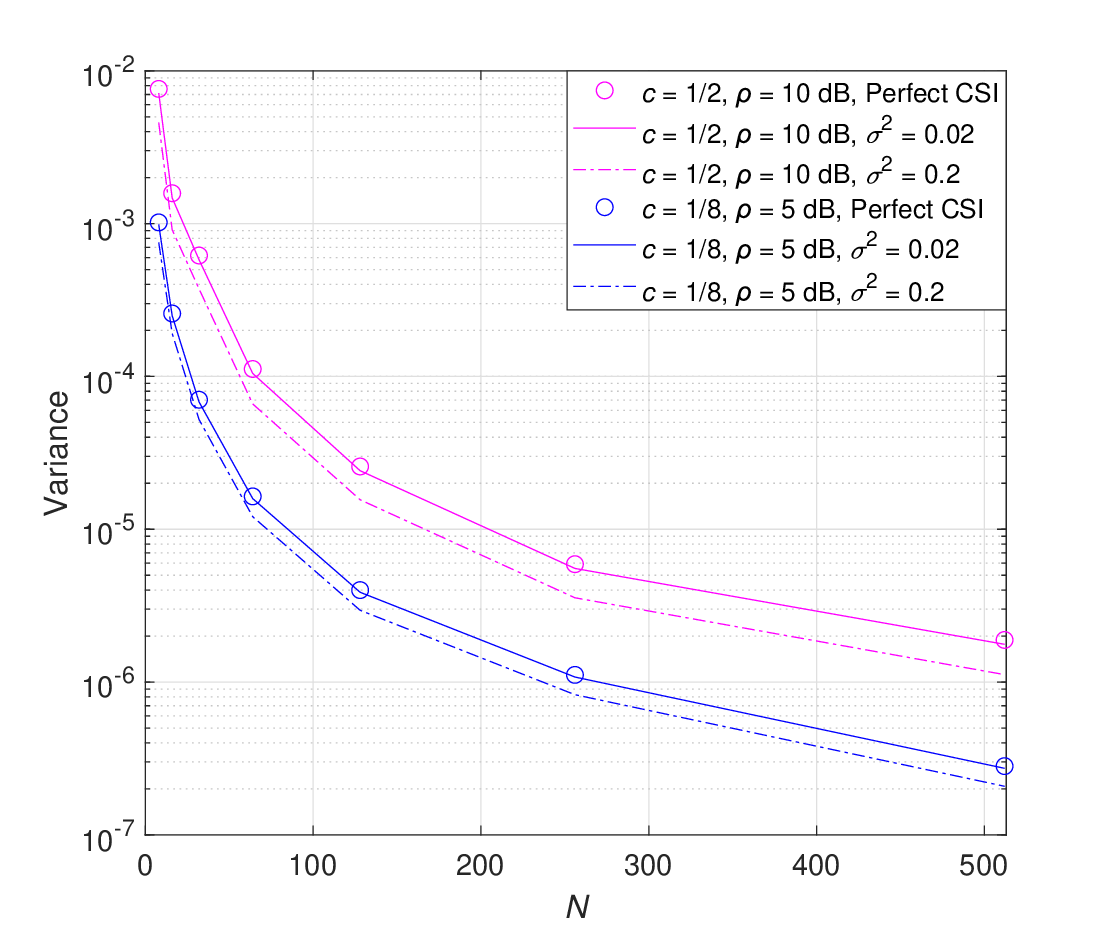}
       \caption{Performance comparison with imperfect CSI.}
     \end{centering}
\end{figure}

\section{Conclusion}
We have derived the closed-form expressions for channel dispersion
in a massive MIMO scenario, which coincide well with the
simulation results. Through using the obtain channel dispersions,
the bound on maximal achievable rate at a finite blocklength can
be expressed by a compact and precious formula in a normal
approximation case which forms an unified framework to explore the
relations among $\epsilon$, $n$, $R$ and the number of spatial
DoF. Based on the new framework, it is proved at the first time
that, although most people think it is intuitively reasonable,
coding along the space domain can also achieve the Shannon
capacity, when the number of antennas goes to infinity. This gives
rise a new methodology, named time space exchanging, to solve the
contradictory problem between data rate and time latency.

%

\appendix
\subsection{Proof of (14)}
Let ${\mathbf{\Lambda }}$ be a diagonal matrix of eigenvalues of
$\bf R$, where ${\mathbf{R}} = {{\mathbf{H}}^H}{\mathbf{H}}$ for
the case of $N > M$. According to the definition of Stieltjes
transform in \cite{Couillet2011}, we have

\begin{equation}
\begin{gathered}
  \sum\limits_{i = 1}^m {\frac{1}{{{{2M} \mathord{\left/
 {\vphantom {{2M} {\rho  + {\lambda _i}}}} \right.
 \kern-\nulldelimiterspace} {\rho  + {\lambda _i}}}}}}  = {\text{tr}}\left[ {{{\left( {{\mathbf{\Lambda }} - z{{\mathbf{I}}_m}} \right)}^{ - 1}}} \right] \hfill \\
  \quad \quad \quad \quad \quad \;\; = m\int {\frac{1}{{\lambda  - z}}} d{F^{\mathbf{R}}}\left( \lambda  \right), \hfill \\
\end{gathered}
\end{equation}
where ${F^{\mathbf{R}}}$ denotes the eigenvalue distribution
function of $\mathbf{R}$ and $z =  - {{2M} \mathord{\left/
 {\vphantom {{2M} \rho }} \right.
 \kern-\nulldelimiterspace} \rho }$. When $m$ grows to infinity, it has been proved that ${F^{\mathbf{R}}}$ converges in distribution (often almost surely so) to some deterministic limit
$F$, which can then be turned into approximative results for
${F^{\mathbf{R}}}$. Furthermore, by using Mar{\v{c}}enko-Pastur
law, the Stieltjes transform has a explicit expression as

\begin{equation}
{\mu _F}\left( z \right) = \frac{{1 - c}}{{2cz}} - \frac{1}{{2c}}
- \frac{{\sqrt {{{\left( {1 - c - z} \right)}^2} - 4cz} }}{{2cz}}
\end{equation}
where $c = {N \mathord{\left/
 {\vphantom {N M}} \right.
 \kern-\nulldelimiterspace} M}$.
Nevertheless, it should be noted that the entries of the matrices
dealt with in this law have different variances with that of the
elements in $\bf{H}$. So, the result of (45) should be scaled.
Based on [32, Lemma 3.2], i.e.,

\begin{equation}
{\mu _{a{\mathbf{R}}}}\left( {az} \right) = \frac{1}{a}{\mu
_{\mathbf{R}}}\left( z \right),
\end{equation}
we obtain

\begin{equation}
{\mu _F}\left( z \right) = a\left( {\frac{{1 - c}}{{2caz}} -
\frac{1}{{2c}} - \frac{{\sqrt {{{\left( {1 - c - az} \right)}^2} -
4caz} }}{{2caz}}} \right),
\end{equation}
where $a = {c \mathord{\left/
 {\vphantom {c M}} \right.
 \kern-\nulldelimiterspace} M}$. Since all the variables in this expression are deterministic, we
say that the expectation approaches to

\begin{equation}
\begin{gathered}
  \mathop {\lim }\limits_{M \to \infty } \mathbb{E}\left\{ {\sum\limits_{i = 1}^m {\frac{1}{{{{2M} \mathord{\left/
 {\vphantom {{2M} {\rho  + {\lambda _i}}}} \right.
 \kern-\nulldelimiterspace} {\rho  + {\lambda _i}}}}}} } \right\} =  \hfill \\
  \quad \quad c\left( {\frac{{1 - c}}{{2caz}} - \frac{1}{{2c}} - \frac{{\sqrt {{{\left( {1 - c - az} \right)}^2} - 4caz} }}{{2caz}}} \right). \hfill \\
\end{gathered}
\end{equation}
Finally, substituting $c = {N \mathord{\left/
 {\vphantom {N M}} \right.
 \kern-\nulldelimiterspace} M}$, $a = {c \mathord{\left/
 {\vphantom {c M}} \right.
 \kern-\nulldelimiterspace} M}$ and $z =  - {{2M} \mathord{\left/
 {\vphantom {{2M} \rho }} \right.
 \kern-\nulldelimiterspace} \rho }$ into (48) yields (14).

On the other hand, when $N \leqslant M$, (45) still holds. So, as
long as we set $c' = {1 \mathord{\left/
 {\vphantom {1 c}} \right.
 \kern-\nulldelimiterspace} c}$ and $a' = {{c'} \mathord{\left/
 {\vphantom {{c'} M}} \right.
 \kern-\nulldelimiterspace} M}$ to scale the matrix, the expectation can then be obtained as

\begin{equation}
\begin{gathered}
  \mathop {\lim }\limits_{M \to \infty } \mathbb{E}\left\{ {\sum\limits_{i = 1}^m {\frac{1}{{{{2M} \mathord{\left/
 {\vphantom {{2M} {\rho  + {\lambda _i}}}} \right.
 \kern-\nulldelimiterspace} {\rho  + {\lambda _i}}}}}} } \right\} =  \hfill \\
  \quad \left( {\frac{{1 - c'}}{{2c'a'z}} - \frac{1}{{2c'}} - \frac{{\sqrt {{{\left( {1 - c' - a'z} \right)}^2} - 4c'a'z} }}{{2c'a'z}}} \right), \hfill \\
\end{gathered}
\end{equation}
where $z =  - {{2M} \mathord{\left/
 {\vphantom {{2M} \rho }} \right.
 \kern-\nulldelimiterspace} \rho }$.

\subsection{Proof of (26)}
If all the eigenvalues are independent of each other, we can
rewrite (21) as
 \begin{equation}
\begin{gathered}
  \mathbb{E}\left\{ {{{\left( {\frac{M}{{2\rho }}} \right)}^2}\sum\limits_{i = 1}^m {\sum\limits_{j = 1}^m {\frac{1}{{{\lambda _i}\left( {{{2M} \mathord{\left/
 {\vphantom {{2M} {\rho  + {\lambda _j}}}} \right.
 \kern-\nulldelimiterspace} {\rho  + {\lambda _j}}}} \right)}}} } } \right\} \hfill \\
  \quad \quad \quad  = {\left( {\frac{M}{{2\rho }}} \right)^2}\mathbb{E}\left\{ {\sum\limits_{i = 1}^m {\frac{1}{{{\lambda _i}}}} } \right\}\mathbb{E}\left\{ {\sum\limits_{j = 1}^m {\frac{1}{{{{2M} \mathord{\left/
 {\vphantom {{2M} {\rho  + {\lambda _j}}}} \right.
 \kern-\nulldelimiterspace} {\rho  + {\lambda _j}}}}}} } \right\}. \hfill \\
\end{gathered}
 \end{equation}
Then, by using the results in (10) and (14), it is easy to obtain
 \begin{equation}
\begin{gathered}
  {\left( {\frac{M}{{2\rho }}} \right)^2}\mathbb{E}\left\{ {\sum\limits_{i = 1}^m {\frac{1}{{{\lambda _i}}}} } \right\}\mathbb{E}\left\{ {\sum\limits_{j = 1}^m {\frac{1}{{{{2M} \mathord{\left/
 {\vphantom {{2M} {\rho  + {\lambda _j}}}} \right.
 \kern-\nulldelimiterspace} {\rho  + {\lambda _j}}}}}} } \right\} \hfill \\
  \quad \quad \quad  = {\left( {\frac{M}{{2\rho }}} \right)^2}\frac{{{N^2}}}{{M - N}}\left[ {\frac{{N - M}}{{4\rho N}} - \frac{1}{2}} \right. \hfill \\
  \quad \quad \quad \quad \quad \quad \left. { + \frac{{\rho \sqrt {{{\left( {M - N + {{2N} \mathord{\left/
 {\vphantom {{2N} \rho }} \right.
 \kern-\nulldelimiterspace} \rho }} \right)}^2} + {{8{N^2}} \mathord{\left/
 {\vphantom {{8{N^2}} \rho }} \right.
 \kern-\nulldelimiterspace} \rho }} }}{{4N}}} \right]. \hfill \\
\end{gathered}
 \end{equation}
However, we all know that the eigenvalues are not independent in a
matrix. They are actually correlated to each other. So, this
expression is not accurate and needs to be emendated. Fortunately,
simulation results show that it has the same trend of variation as
that of $\text{G2}$ and their difference can be corrected by a
constant parameter. Thus, we finally obtain

\begin{equation}
\begin{gathered}
  {\text{G2}} = \zeta {\left( {\frac{M}{{2\rho }}} \right)^2}\frac{{{N^2}}}{{M - N}}\left[ {\frac{{N - M}}{{4\rho N}} - \frac{1}{2}} \right. \hfill \\
  \quad \quad \quad \quad \quad \quad \left. { + \frac{{\rho \sqrt {{{\left( {M - N + {{2N} \mathord{\left/
 {\vphantom {{2N} \rho }} \right.
 \kern-\nulldelimiterspace} \rho }} \right)}^2} + {{8{N^2}} \mathord{\left/
 {\vphantom {{8{N^2}} \rho }} \right.
 \kern-\nulldelimiterspace} \rho }} }}{{4N}}} \right], \hfill \\
\end{gathered}
 \end{equation}
where $\zeta  = {1 \mathord{\left/
 {\vphantom {1 {\left( {\psi N} \right)}}} \right.
 \kern-\nulldelimiterspace} {\left( {\psi N} \right)}}$ denotes
 the emendation parameter and $\psi $ is a real number which has different values for different $\rho$.

\subsection{Proof of (34)}
At first, we consider $M>N$ and obtain
\begin{equation}
\begin{gathered}
  \mathbb{E}\left[ {V\left( {\mathbf{H}} \right)} \right] \approx m - \frac{m}{{{\rho ^2}}} \cdot \frac{{{M^3}}}{{{{\left( {M - m} \right)}^3} - \left( {M - m} \right)}} \hfill \\
  \quad \quad \quad \quad  = m\left( {1 - \frac{1}{{{\rho ^2}}} \cdot \frac{{{M^3}}}{{{{\left( {M - m} \right)}^3} - \left( {M - m} \right)}}} \right) \hfill \\
  \quad \quad \quad \quad  = m\left( {1 - \frac{1}{{{\rho ^2}}} \cdot {\text{G}}5} \right). \hfill \\
\end{gathered}
\end{equation}
In practical systems, the number of transmit antennas is usually
set to be an integer multiple of the receive antennas, such as 5G
downlink, to facilitate the use of multi-user MIMO or support
coordinated multiple points (CoMP) transmissions. In this case,
$M$ is an even number and $m$ ranges from $1$ to $M/2$, i.e., $1
\leqslant m \leqslant {M \mathord{\left/
 {\vphantom {M 2}} \right.
 \kern-\nulldelimiterspace} 2}$. Substituting the extreme values into ${\text{G}}5$ and assuming $M>2$, we obtain

\begin{equation}
\begin{gathered}
  \frac{{{M^3}}}{{{{\left( {M - m} \right)}^3} - \left( {M - m} \right)}} < \frac{{{M^3}}}{{{{\left( {M - \frac{M}{2}} \right)}^3} - \left( {M - 1} \right)}} \hfill \\
  \quad \quad \quad \quad \quad \quad \quad \quad  = \frac{{8{M^2}}}{{{M^2} - 8 + \frac{8}{M}}} < 13. \hfill \\
\end{gathered}
\end{equation}
In the high SNR regime, $\rho$ is of course larger than $\sqrt
{13}$. So,
\begin{equation}
\frac{1}{{{\rho ^2}}} \cdot {\text{G}}5 < 1
\end{equation}
holds obviously. When we further increase $\rho$ to a very large
number, the term in (53) diminishes completely and (34) is
obtained then.

For the case of $M<N$, the bound of $\rho$ to make ${\text{G}}5$
be smaller than one can be found by using the same way, but we
would not repeat it here, due to the limited room.


\begin{thebibliography}{99}

\bibitem{Sachs2018}
J.~Sachs, G.~Wikstrom, T.~Dudda, \emph{~et~al.}, ``5G radio
network design for ultra-reliable low-latency communication,''
\emph{IEEE Network}, vol. 32, no. 2, pp. 24--31, 2018.

\bibitem{Liu2022}
S.~Liu, C.~Zheng, Y.~Huang, and T.~Q.~S.\ Quek, ``Distributed
reinforcement learning for privacy-preserving dynamic edge
caching,'' \emph{IEEE J. Sel. Areas Commun.}, vol. 40, no. 3, pp.
749--760, 2022.

\bibitem{3GPP2017}
3rd Generation Partnership Project (3GPP), \emph{Evolved Universal
Terrestrial Radio Access (E-UTRA); Radio Resource Control (RRC);
Protocol specification}, Technical Specification (TS) 36.331, 04
2017.

\bibitem{Sutton2019}
G.~J.\ Sutton, J.~Zeng, R.~P.\ Liu, \emph{~et~al.}, ``Enabling
technologies for ultra-reliable and low latency communications:
From PHY and MAC layer perspectives,'' \emph{IEEE Commun. Surv.
Tutor.}, vol. 21, no. 3, pp. 2488--2524, 2019.

\bibitem{Xu2023}
W. Xu, Y. Huang, W. Wang, F. Zhu, and X. Ji, ``Toward ubiquitous
and intelligent 6G networks: From architecture to technology,''
\emph{Science China Information Sciences}, vol. 66, no. 3, pp:
5-6, Mar. 2023.



\bibitem{Saad2019}
W. Saad, M. Bennis, and M. Chen, ``A vision of 6G wireless
systems: Applications, trends, technologies, and open research
problems,'' \emph{IEEE Network}, vol. 34, no. 3, pp. 134--142,
2019.

\bibitem{XuWei}
W. Xu, Z. Yang, D. W.-K. Ng, M. Levorato, Y. C. Eldar, and M.
Debbah, ``Edge learning for B5G networks with distributed signal
processing: Semantic communication, edge computing, and wireless
sensing,'' \emph{IEEE J. Sel. Topics Signal Process.}, vol. 17,
no. 1, pp. 9--39, Jan. 2023.


\bibitem{You2021}
X. You, C. Wang, J. Huang, \emph{~et~al.}, ``Towards 6G wireless
communication networks: Vision, enabling technologies, and new
paradigm shifts,'' \emph{Science China Information Sciences}, vol.
64, no. 1, pp. 1--74, 2021.


\bibitem{Zorzi1} M. Giordani, M. Polese, M.
Mezzavilla, S. Rangan, and M. Zorzi, ''Toward 6G Networks: Use
Cases and Technologies,'' \emph{IEEE Commun. Mag.}, vol. 58, no.
3, pp. 55--61, Mar. 2020.


\bibitem{You2023}
X. You, ``6G extreme connectivity via exploring spatiotemporal
exchangeability,'' \emph{Science China Information Sciences}, vol.
66, no. 3, pp. 94--96, 2023.


\bibitem{Liu2021}
Y. Liu, Y. Deng, M. Elkashlan, A. Nallanathan, and G. K.
Karagiannidis, ``Analyzing grant-free access for URLLC service,''
\emph{IEEE J. Sel. Areas Commun.}, vol. 39, no. 3, pp. 741--755,
Mar. 2021.


\bibitem{She2019}
C. She, Y. Duan, G. Zhao, T. Q. S. Quek, Y. Li and B. Vucetic,
``Cross-layer design for mission-critical IoT in mobile edge
computing systems,'' \emph{IEEE Internet Things J.}, vol. 6, no.
6, pp. 9360--9374, Dec. 2019.

\bibitem{She2018}
C. She, Z. Chen, C. Yang, T. Q. S. Quek, Y. Li and B. Vucetic,
``Improving network availability of ultra-reliable and low-latency
communications with multi-connectivity,'' \emph{IEEE Trans.
Commun.}, vol. 66, no. 11, pp. 5482--5496, Nov. 2018.

\bibitem{Bui2017}
N. Bui, M. Cesana, S. A. Hosseini, Q. Liao, I. Malanchini and J.
Widmer, ``A survey of anticipatory mobile networking:
Context-based classification, prediction methodologies, and
optimization techniques,'' \emph{IEEE Commun. Surv. Tutor.}, vol.
19, no. 3, pp. 1790--1821, Q3 2017.

\bibitem{Shannon1968}
C. E. Shannon, ``A mathematical theory of communication,''
\emph{Bell System Technical Journal}, vol. 27, pp. 379--423, 1948.

\bibitem{Gallager1968}
R. G. Gallager, \emph{Information Theory and Reliable
Communication}, New York, USA, Wiley, 1968.

\bibitem{You2020}
X. You, ``Shannon theory and future 6G's technique potentials,''
\emph{Scientia Sinica Informationis}, vol. 50, no. 9, pp.
1377--1394, 2020.

\bibitem{Polyanskiy2010}
Y. Polyanskiy, H. V. Poor, and S. Verdu, ``Channel coding rate in
the finite blocklength regime,'' \emph{IEEE Trans. Inf. Theory},
vol. 56, no. 5, pp. 2307--2359, May 2010.

\bibitem{Polyanskiy2011}
Y. Polyanskiy and S. Verdu, ``Scalar coherent fading channel:
Dispersion analysis,'' in \emph{Proceedings of the IEEE
International Symposium on Information Theory}, pp. 2959--2963,
Aug. 2011.

\bibitem{Yang2012}
W. Yang, G. Durisi, T. Koch, and Y. Polyanskiy, ``Diversity versus
channel knowledge at finite block-length,'' in Proc. \emph{IEEE
Information Theory Workshop}, pp. 572--576, Sept. 2012.

\bibitem{Lapidoth2005}
A. Lapidoth, ``Diversity versus channel knowledge at finite
block-length,'' \emph{IEEE Trans. Inf. Theory}, vol. 51, no. 2,
pp. 437--446, Feb. 2005.


\bibitem{Hochwald2000}
B. M. Hochwald and T. L. Marzetta, ``Unitary space-time modulation
for multiple-antenna communications in Rayleigh flat fading,''
\emph{IEEE Trans. Inf. Theory}, vol. 46, no. 2, pp. 543--564, Mar.
2000.

\bibitem{Zheng2002}
L. Zheng and D. Tse, ``Communication on the Grassmann manifold: A
geometric approach to the noncoherent multiple-antenna channel,''
\emph{IEEE Trans. Inf. Theory}, vol. 48, no. 2, pp. 359--383, Feb.
2002.


\bibitem{Durisi2011}
G. Durisi and H. Bolcskei, ``High-SNR capacity of wireless
communication channels in the noncoherent setting: A primer,''
\emph{Int. J. Electron. Commun. (AEU)}, vol. 65, no. 8, pp.
707--712, Aug. 2011.



\bibitem{Lancho2020}
A. Lancho, T. Koch, and G. Durisi, ``On single-antenna Rayleigh
block-fading channels at finite blocklength,'' \emph{IEEE Trans.
Inf. Theory}, vol. 66, no. 1, pp. 496--519, Jan. 2020.


\bibitem{Zorzi3} B. Makki, T. Svensson, and M.
Zorzi, ``Finite block-length analysis of spectrum sharing networks
using rate adaptation,'' \emph{IEEE Trans. Commun.}, vol. 63, no.
8, pp. 2823--2835, Aug. 2015.


\bibitem{Telatar1999}
E. Telatar, ``Capacity of multi-antenna Gaussian channels,''
\emph{European Trans. Telecommun.}, vol. 10, no. 6, pp. 585--595,
Dec. 1999.

\bibitem{Zheng2003}
L. Zheng and D. Tse, ``Diversity and multiplexing: A fundamental
tradeoff in multiple-antenna channels,'' \emph{IEEE Trans. Inf.
Theory}, vol. 49, no. 5, pp. 1073--1096, May 2003.

\bibitem{Yang2014}
W. Yang, G. Durisi, T. Koch, and Y. Polyanskiy, ``Quasi-static
multiple-antenna fading channels at finite blocklength,''
\emph{IEEE Trans. Inf. Theory}, vol. 60, no. 7, pp. 4232--4265,
Jul. 2014.

\bibitem{Zorzi2} B. Makki, T. Svensson, G. Caire,
and M. Zorzi, ``Fast HARQ over finite blocklength codes: A
technique for low-latency reliable communication,'' \emph{IEEE
Trans. Wireless Commun.}, vol. 18, no. 1, pp. 194--209, Jan. 2019.


\bibitem{Collins2016}
A. Collins and Y. Polyanskiy, ``Dispersion of the coherent MIMO
block-fading channel,'' in Proc. \emph{IEEE International
Symposium on Information Theory (ISIT)}, 2016, Barcelona, Spain,
pp. 572--576.

\bibitem{Ostman} J. Ostman, A. Lancho, G. Durisi,
and L. Sanguinetti, ``URLLC with massive MIMO: Analysis and design
at finite blocklength,'' \emph{IEEE Trans. Wireless Commun.}, vol.
20, no. 10, pp. 6387--6401, Oct. 2021.


\bibitem{Lancho}
A. Lancho, J. Ostman, G. Durisi, and L. Sanguinetti, ``A
finite-blocklength analysis for URLLC with massive MIMO,'' in
Proc. \emph{IEEE International Conference on Communications},
Montreal, QC, Canada, Jun. 2021.


\bibitem{Kislal}
A. Kislal, A. Lancho, G. Durisi, and E. Strom, ``Efficient
evaluation of the error probability for pilot-assisted URLLC with
Massive MIMO,'' \emph{https://arxiv.org/abs/2211.02385}.


\bibitem{Potter}
C. Potter, K. Kosbar, and A. Panagos, ``On achievable rates for
MIMO systems with imperfect channel state information in the
finite length regime,'' \emph{IEEE Trans. Commun.}, vol. 61, no.
7, pp. 2772--2781, Jul. 2013.


\bibitem{Marzetta2010}
T. L. Marzetta, ``Noncooperative cellular wireless with unlimited
numbers of base station antennas,'' \emph{IEEE Trans. Wireless
Commun.}, vol. 9, no. 11, pp. 3590--3600, Nov. 2010.

\bibitem{Abbe2013}
E. Abbe, S.-L. Huang, and I. E. Telatar, ``Proof of the outage
probability conjecture for MISO channels,'' \emph{IEEE Trans. Inf.
Theory}, vol. 59, no. 5, pp. 2596--2602, May 2013.

\bibitem{Lozano2003}
A. Lozano, A. M. Tulino, and S. Verdu, ``Multiple-antenna capacity
in the low-power regime,'' \emph{IEEE Trans. Inf. Theory}, vol.
49, no. 10, pp. 2527--2544, Oct. 2003.

\bibitem{Jungnickel2002}
V. Jungnickel, T. Haustein, E. Jorswieck, and C. von Helmolt, ``On
linear pre-processing in multi-antenna systems,'' in \emph{Proc.
IEEE GLOBECOM}, Taiwan, China, pp. 1012--1016, Nov. 2002.


\bibitem{Tulino2004}
A. M. Tulino and S. verdu, \emph{Random Matrix Theory and Wireless
Communications}, Now Foundations and Trends, 2004.


\bibitem{Couillet2011}
R. Couillet and M. Debbah, \emph{Random Matrix Methods for
Wireless Communications}, Cambridge University Press, 2011.


\bibitem{Tse2005} D. Tse and P. Viswanath,
\emph{Fundamentals of Wireless Communication}, Cambridge
University Press, 2005.

\bibitem{Jayaweera}
N. Jayaweera, D. Marasinghe, N. Rajatheva, and M. Latva-aho,
``Factory automation: resource allocation of an elevated LiDAR
system with URLLC requirements," in \emph{Proc. 2nd 6G Wireless
Summit (6G SUMMIT)}, Levi, Finland, Mar. 2020.


\bibitem{Yang}
B. Yang, K. Letaief, R. Cheng, and Z. Cao, ``Channel estimation
for OFDM transmission in multipath fading channels based on
parametric channel modeling,'' \emph{IEEE Trans. Commun.}, vol.
49, no. 3, pp. 467--479, Mar. 2001.



\end{thebibliography}

\balance

\end{document}